\documentclass[a4paper,11pt]{article}
\usepackage{jheppub}
\usepackage{epsfig,epsf}
\usepackage{mathrsfs}
\usepackage{graphicx}
\usepackage{latexsym}
\usepackage{amsmath}
\usepackage{amssymb}
\usepackage{textcomp}
\usepackage{amsbsy}
\usepackage{tabularx}
\usepackage{array}
\DeclareGraphicsExtensions{.png,.eps,.pdf}
 \usepackage[normalem]{ulem}
 
\definecolor{darkgreen}{cmyk}{1,0,1,0.4}

\long\def\/*#1*/{}
\renewcommand{\baselinestretch}{1.5}

\def\beq{\begin{equation}}
\def\eeq{\end{equation}}
\def\barr{\begin{array}}
\def\earr{\end{array}}
\def\dis{\displaystyle}
\def\tev{\, {\rm TeV}}
\def\gev{\, {\rm GeV}}

\def\lapp{\mathrel{\rlap{\raise.5ex\hbox{$<$}}
                    {\lower.5ex\hbox{$\sim$}}}}
\def\gapp{\mathrel{\rlap{\raise.5ex\hbox{$>$}}
                    {\lower.5ex\hbox{$\sim$}}}}
\def\sup{{\cal S}_{\rm up}}
\def\sdn{{\cal S}_{\rm dn}}
\def\ft{\widetilde f}
\def\gym{g_{\mbox{\lower.25ex\hbox{\tiny YM}}}}

\title{
Bulk gauge and matter fields in nested warping: II. Symmetry Breaking and phenomenological consequences}

\author[1]{Mathew Thomas Arun\note{Corresponding author.},}
\author{Debajyoti Choudhury}
\emailAdd{thomas.mathewarun@gmail.com}

\emailAdd{debajyoti.choudhury@gmail.com} \affiliation[a]{Department of
  Physics and Astrophysics, University of Delhi, Delhi 110007, India.}
\abstract {Generalizing the Randall-Sundrum scenario to higher
  dimensions with nested warpings has been shown to avoid the
  constraints besetting the former.  In the first paper of this series
  [JHEP 1509 (2015) 202], the Standard Model gauge and fermion fields
  were extended into such a six-dimensional bulk and the construction
  was shown to have several interesting and welcome features.  In this
  paper, we discuss the electroweak symmetry breaking, presenting a
  novel Higgs localization mechanism that leads to interesting
  phenomenology in the Higgs sector. Localizing the Higgs modifies the
  $Z_{\mu}$ and $W_{\mu}$ boson wavefunctions, which leads to tree
  level changes in the oblique parameters. Using these as well as the
  correction to low-energy four-Fermi operators, we derive the
  constraints on our model and also discuss the gauge coupling
  evolution therein.  Amusingly, the model can naturally incorporate a
  Higgs resonance in the 700--800 $\gev$ range.  }
\begin{document}
\maketitle
\flushbottom

\newpage
\section{Introduction}
It has long been recognized that theories
defined in dimensions larger than four may provide geometric
resolutions to some of the quandaries faced by the Standard Model.
Amidst diverse theoretical constructs addressing such issues, have
been efforts~\cite{Rubakov:1983bb, Visser:1985qm,Antoniadis:1990ew,
Gogberashvili:1998vx, Gogberashvili:1998iu,
Randall:1999ee,Randall:1999vf,ArkaniHamed:1998rs,
Antoniadis:1998ig} to intertwine gravity with low
energy phenomenology.  The Randall-Sundrum (RS)
model\cite{Randall:1999ee} and its extensions comprise 
one such set of endeavours.
Unlike in the ADD scenario\cite{ArkaniHamed:1998rs}, wherein the
hierarchy is sought to be explained by introducing a large volume in
the extra dimensions, in the RS model it is done by postulating a
nonfactorizable geometry with an exponential warping between two flat
3-branes. While we are located on the TeV brane (wherein the natural
scale of the theory, viz. $M_{\rm Pl}$, is warped down to and
perceived as the TeV scale), the other (Planck) brane remains hidden.

A more interesting (from the particle physics point of
view) model is constructed by a minimal extension of this RS
model with gauge bosons in the bulk and fermions
stuck to the brane \cite{Pomarol:1999ad, Davoudiasl:1999tf}.
Such forays into the bulk come at a cost, though. For example, the
gauge boson KK-excitations couple to fermion bilinears almost
universally and with a strength approximately eight times as large as
that of the zero mode.  This result, in conjunction with the global
fits on the four Fermi operator \cite{Rizzo:1999br}, demand that the
first excited mode mass must be $> 23 \tev$.  To appreciate this
constraint, it is useful to reexpress it in terms of the model
parameters, viz. the fundamental five-dimensional mass $M_{5D}$, the
radius of compactification $R_y$ and the exponential ($e^{-c|y|}$)
warping parameter $c$, whereby it translates to $c/(R_y M_{5D}) >
4.5$.  On the other hand, the very applicability of semi-classical
arguments, on which the entire RS construction hinges, calls for this
combination to be $\lapp 0.1$, thus calling into question the
trustworthiness of this approach.

On allowing the fermions too to enter the bulk, it was shown
\cite{Chang:1999nh} that the coupling of the fermion zero-mode to the
first KK gauge boson could be suppressed significantly, thereby
relaxing the constraints from the four-Fermi operator.  On the other
hand, since the mass hierarchy problem can be solved only by using a
TeV-brane localized Higgs field, the latter's gauge coupling deforms
the boundary conditions on the gauge bosons.  The consequent
distortion in the profile of the lowest gauge boson, results in tree
level corrections to the electroweak oblique parameters
\cite{Csaki:2002gy}.  Consistency with the precision data now demands
that the first KK-mode for the gauge boson be heavier than $27\tev$
(or, equivalently, $c R_y^{-1}e^{- c \pi} > 11 \tev$), thereby
resurrecting the problem in a different guise.  This, though, could be
cured, albeit at the expense of introducing a custodial symmetry in
the bulk\cite{Agashe:2003zs}.  The enhanced gauge structure preserves the
isospin symmetry and thus softens the constraint on the $T$-parameter.
Similarly, localizing the light fermions near the Planck-brane
controls the $S$-parameter, such that the precision test data fits are
satisfied by a KK gauge boson with mass of a few TeVs.

On a track parallel to this, emerged several attempts in creating
models in ($5+1$)-dimensions.  While the flat space
variants~\cite{Dobrescu:2001ae,Appelquist:2001mj,
  Burdman:2006gy,Dobrescu:2007xf,Freitas:2007rh,Choudhury:2011jk,Dobrescu:2007ec,
  Cacciapaglia:2009pa} did consider bulk matter fields so as to
address some of the lacunae of the SM, those with non-factorizable
geometries ~\cite{Shaposhnikov,Nelson,Cohen,Giovannini,
  Kanti,ChenF,Gogberashvili:2003xa,
  Gogberashvili:2003ys,Gogberashvili:2007gg,Choudhury:2006nj}
typically restricted themselves to discussion of the hierarchy and/or
cosmological issues. Although seemingly modest in their aspiration,
the latter set of constructions have recently gained relevance in the
context of negative results achieved by both the
ATLAS~\cite{ATLAS2015} for RS graviton resonances. While reasonable values for
the ratio of the five-dimensional curvature and the fundamental mass
scale would predict that the mass of the first KK-graviton be a few
times larger than that of the Higgs, the current lower limit of $\sim
2.66\tev$ (at 95\% C.L.) is already causing some tension for the
scenario. In Ref.\cite{Arun:2014dga}, though, it was demonstrated
that, in the event of nested warping in a 6D scenario, the graviton
modes comfortably evade the current bounds from the LHC.  Furthermore,
as Ref.\cite{Arun:2015ubr} points out, not only is the allowed
parameter space of the model quite extensive and can be probed well in
the current run of LHC, it also admits an explanation of the recently
reported anomaly\cite{atlas_750, cms_750} at $m_{\gamma \gamma} \sim
750 \gev$.

It is, thus, interesting
to consider the possibility of allowing the SM fields into the bulk of
such a nested warping scenario, and we had developed this formalism in
ref.~\cite{Arun:2015kva}, hereafter referred to as Paper I. The
construction has several striking features.  The most notable is that,
apart from offering an ``explanation'' of the number of fermion
generations, it essentially ``localizes'' part of the fermions onto a
4-brane. This has the immediate consequence that whereas the gauge
bosons (and, of course, the graviton) have a ``tower of KK-towers'', 
for the fermions one of the towers is missing. {This would have
striking ramifications in collider searches, both in terms of the
observed low-energy spectrum as well as in the decay patterns (and,
hence, in the signature topologies). Furthermore, the ``missing''
fermionic states would leave imprint in both corrections to
observables as well as in engendering rare processes. In the present
work, we examine some of such phenomenological consequences. 

The rest of this paper is structured as follows.  We start out with a
brief recapitulation of the scenario augmented by a discussion of
fermion mixing (an aspect that was glossed over
earlier). Subsequently, in Sec.\ref{sec:higgs}, we consider the Higgs
sector in detail and present the Higgs spectrum for the particular
localization that we employ. This is followed, in Sec.\ref{sec:l_eff},
by the derivation of an effective Lagrangian that allows us to
  reliably calculate four-Fermi operators as well as the electroweak
  precision observables. Utilizing this, in Sec.\ref{sec:chisq}, to
  constrain the parameter space, we next investigate (in
  Sec.\ref{sec:rg}) the renormalization group flow of the gauge
  couplings, which allows us examine the nature of gauge
  unification. Finally, we summarise in Sec.\ref{sec:conclusion}.


\section{Gauge and fermion fields}
\label{recap}
We consider a six-dimensional space-time compactified down to 
four dimensions with a  $Z_2$ orbifolding in each of the two extra dimensions,
viz.  $M^{1,5}\rightarrow[M^{1,3}\times S^1/Z_2]\times S^1/Z_2$. A successive
(nested) warping is assumed leaving the four-dimensional space to be flat. 
In other words, the line element is of the form~\cite{Choudhury:2006nj}
\begin{equation}
ds^2= b^2(x_5) \, [a^2(x_4)\eta_{\mu\nu}dx^{\mu}dx^{\nu}+R_y^2dx_4^2]
     +r_z^2dx_5^2 \ ,
\label{metric}
\end{equation}  
where the compact directions are represented by the dimensionless
coordinates $ x_{4,5}\in [0,\pi]$ with $R_y$ and $r_z$ being the
corresponding moduli. The background geometry is given by 
the six-dimensional Einstein-Hilbert action (with a natural scale $M_6$)
and a negative (six-dimensional) cosmological constant $\Lambda_6$ 
yielding\cite{Choudhury:2006nj}
\begin{equation}
\barr{rcl c rclcrcl}
a(x_4)& = & e^{-c|x_4|} 
     &\qquad &
     c & = & \dis \frac{R_y k}{r_z\cosh{k\pi}} 
       & \equiv & \dis \frac{\aleph \, k}{\cosh(k \pi)}
\\[1ex]
b(x_5)& = & \dis \frac{\cosh{(k x_5)}}{\cosh{(k\pi)}} 
     &\qquad\qquad& 
     k& = & \dis r_z\sqrt{\frac{-\Lambda_6}{10 M_6^4}} 
      & \equiv  & \dis \epsilon \, r_z \,M_6 \ .
\earr
                \label{RS6_eqns}
\end{equation}
The difference in scale between the Planck brane and the $\tev$ brane,
where the Higgs is localized, sets a measure for $w$, the extent of the
  hierarchy. Typically, $w$ ranges from $e^{-c \pi}$ to $e^{-c \pi} \,
  {\rm sech} k \pi$, with the exact value depending on the details of
  the Higgs localization.

Clearly, we can consistently neglect quantum corrections to the bulk
gravity action (necessary for the validity of the semi classical
treatment) only if the bulk curvature is significantly smaller than
the fundamental scale $M_6$, or in other words if $\epsilon \lapp
0.1$.  On the other hand, the requirement of not reintroducing a large
hierarchy requires that $\aleph$ (the ratio of the two moduli) should
not be too large. This, along with the phenomenological requirement of
$w \sim 10^{-16}$ (or even an order of magnitude or two larger) forces
the theory into one of two branches, namely $(i)$ $c \sim {\cal
  O}(10), k \lapp 1$ or $(ii)$ $k \sim {\cal O}(10)$ and a negligibly
small $c$ \cite{Choudhury:2006nj}. While each branch has its merits,
the second one results in considerably enhanced couplings for the
KK-gravitons\cite{Arun:2014dga}. Furthermore, once gauge fields are
allowed to go into the bulk, their KK-excitations, for this branch of
the theory, are bestowed with too large a coupling to admit
perturbation theory\cite{Arun:2015kva}.  Consequently, we shall
concentrate on the first branch alone.

We start our review of the SM fields with the gauge sector, which,
along with the fermions, percolates fully into the bulk. The kinetic term,
for a theory with unbroken symmetry, is thus
given by
\begin{equation}
\barr{rcl}
{\cal L}& =& \dis \frac{-1}{4}\sqrt{-g}F_{M N}F^{M N} + {\cal L}_{gf}
\\[2ex]
{\cal L}_{gf} & = & \dis \frac{-\sqrt{-g}}{2 \zeta} \,
     \left[ g^{\mu\nu} \left\{\partial_\mu A_\nu 
                 - \frac{\zeta}{2} \, \left(\Gamma^{4}_{\mu \nu} A_4 
                                      +\Gamma^{5}_{\mu \nu} A_5\right) \right\}
            + \zeta \, (g^{44} D_4 A_4 + g^{55} D_5 A_5) \right]^2
\\[2ex]
& = & \dis
\frac{-R_y r_z b}{2 \zeta} \left[ \eta^{\mu\nu} \partial_{\mu}A_{\nu} 
       + \frac{\zeta}{b} \left( \partial_4 \frac{a^2 b A_4}{R_y^2} 
                              + \partial_5 \frac{a^2 b^3 A_5}{r_z^2} \right)
           \right]^2 \ ,
\earr
\end{equation}
where the choice of the gauge-fixing term (a curved-space analog of 
the generalized $R_\zeta$ gauge) eliminates the cumbersome kinetic mixing
terms between $A_{\mu}$ and $A_{4,5}$. Writing $A_{\mu}$ in terms of the 
KK modes, viz.
\[
 A_\kappa = 
\frac{1}{\sqrt{R_y r_z}} \sum_{n,p} 
A^{(n,p)}_\kappa(x^{\mu}) \, \eta_{n,p}(x_4) \, \chi_p(x_5) 
\]
with $\eta_{n,p}$ and $\chi_p$ normalized as
\[
\dis \int_{}^{} b(x_5) \chi_{p_1}\chi_{p_2} dx_5  =  \delta_{p_1,p_2}  \ ,
\qquad \quad
\int_{}^{} \eta_{n_1,p}\eta_{n_2,p} dx_4  = \delta_{n_1,n_2} \ ,
\]
the solutions for the modes are 
\beq
\barr{rcl}
\chi_p(x_5) & = & \dis \frac{1}{B} {\rm sech}^{3/2}(k x_5) \,
      \left( c_1 \, P_{\nu_p}^{3/2}(\tanh k x_5) + 
             c_2 \, Q_{\nu_p}^{3/2}(\tanh k x_5)\right)
\\[2ex]
\eta_{n,p}(x_4) & = & \dis 
    \frac{e^{c|x_4|}}{N} \Big( J_{\nu_n}(y_n) + c_{np} Y_{\nu_n}(y_n)\Big)
\\[2ex]
y_n & \equiv & \dis m_{np}\frac{r_z}{k} e^{c|x_4|} \, \cosh(k \pi) = 
                    m_{np} \, \frac{R_y}{c} \, e^{c|x_4|} 
\\[2ex]
\nu_n & = & \dis \sqrt{1+\frac{r_z^2}{k} \, m_p^2 \, \cosh^2(k \pi)}  
\\[2ex]
\nu_p & = & \dis \frac{-1}{2} + \nu_n \ .
\earr
    \label{gauge_soln}
\eeq
Before we impose the boundary conditions on the $\chi$'s and the $\eta$'s
(and, thereby, compute the spectrum), let us remind ourselves that the
electroweak symmetry has, of course, to be broken spontaneously.
While this could, in principle, be done with a bulk Higgs field, such
a course of action would imply that the Higgs mass (or the vacuum
expectation value) would assume the natural scale, namely $M_{6}$, and
the hierarchy problem would resurface. This is exactly analogous to the case 
of the corresponding five-dimensional scenario. A way out would be to 
confine the Higgs to a brane wherein the perceived scale is naturally low. 
In the present case, it could be 3-brane located at $(x_4 = \pi, x_5 = 0)$, 
or, more generally, the 4-brane at $x_4 = \pi$. As has been pointed out in 
Ref.\cite{Arun:2015kva}, the first course of action leads to a equation 
of motion for the gauge bosons that does not let itself to a closed-form 
solution commensurate with the boundary conditions.
To this end, we consider a theory with an explicit cutoff\footnote{It is at this scale that the compactified direction $x_4$ would reveal itself and a four-dimensional description would no longer be tenable.
 $\lapp R_y^{-1}$ and described by a Higgs Lagrangian of the form}
\beq
\barr{rcl}
{\cal L}_h & = & \dis \delta(x_4-\pi) \sqrt{-g_5}
    \Big( g^{\mu \nu}D_{\mu} \phi(x^{\bar M})^{\dagger} D_{\nu} \phi(x^{\bar M}) + \aleph^{-2} g^{55}D_{5} \phi(x^{\bar M})^{\dagger} D_{5}
+ V(\phi) \Big) \ ,
\\[2ex]
D_{\mu} & = & \dis \partial_{\mu} -i \, \gym\, A_{\mu}(x_{\nu},x_4,x_5)
\earr
    \label{4brane_h_lagr}
\eeq
where the barred indices ($\bar M$
etc.) run over the coordinates ($0,1,2,3,5$) relevant to this brane. 
Note that the form of the Lagrangian is slightly different from that 
proposed in Ref.\cite{Arun:2015kva}. In particular, the factor $\aleph^{-2}$ 
ensures that the natural scale of the theory is $R_y^{-1}$ and not $r_z^{-1}$ 
(which is larger than the cutoff). While the form above is seemingly 
inconsistent with the full five-dimensional Lorentz invariance, this 
is not of concern here. In fact, the very presence of the $x_5$-dependent 
brane tension $V_2(x_5)$~\cite{Choudhury:2006nj,Arun:2015kva} has already 
destroyed part of the symmetry
leaving behind a manifest four-dimensional  Lorentz invariance. 
$V(\phi)$ is a 
potential admitting a nontrivial vacuum and, thus, a brane-localized 
mass term for the gauge boson. The solution to the corresponding 
gauge equation of motion is still rather complicated, but can be simplified 
substantially if $V(\phi)$ is such that the scalar equation of motion admits
a $x_5$-dependent profile of the form 
$\left\langle \phi(x_5) \right\rangle \propto v / \sqrt{b(x_5)}$ with 
$v$ being the (constant) vacuum expectation value as mentioned in Ref.\cite{Arun:2015kva}.
Postponing discussions about the form of the $V(\phi)$ needed, 
we assume that the profile is indeed so. This would, then, introduce 
a brane-localized gauge field mass term of the form
\[
 {\cal L}_m = 
 \frac{\sqrt{-g_5}}{2} \, \widetilde M^2(x_5) \, g_5^{\mu \nu} A_{\mu}A_{\nu} 
     \delta(x_4-\pi)
\]
with $\widetilde M = m / \sqrt{b(x_5)}$ where $m \propto \gym \, v$. The consequent boundary conditions are
\[
\chi_p'|_{x_5=0} = 0 = \chi_p'|_{x_5=\pi} 
\]
and 
\begin{equation}
\label{bcy1}
\eta'_{n,p}|_{x_4=0} = 0 \ , 
\quad {\rm and} \quad
\eta'_{n,p}|_{x_4=\pi} =  m^2 \, R_y^2 \, \eta_{n,p}(\pi) \ .
\end{equation}
For $m_{p=0} = 0 $, we have, for the modes $\eta_{n0}$,
\[
J_0(e^{-c\pi}\alpha_{n0})\Big( 2 c \alpha_{n0} Y_0(\alpha_{n0}) + R_y^2m^2Y_1(\alpha_{n0})\Big) =
Y_0(e^{-c\pi}\alpha_{n0})\Big( 2 c \alpha_{n0} J_0(\alpha_{n0}) + R_y^2m^2J_1(\alpha_{n0})\Big) ,
\] 
where, as before, $\alpha_{n0} \equiv m_{n0} \, R_y e^{c\pi} / c$.  Since the lightest 
mass mode is to be identified with the
$W/Z$ bosons, we have $\alpha_{00} \sim m_{00} R_y \, e^{c \pi}/ c \ll 1$ (as $c
\sim 10$). Expanding the Bessel functions, we obtain
\beq
 m^{2}_{00} \approx \frac{1}{2 \pi} m^2 \, e^{-2c \pi} \ .
\label{gaugebosonmass}
\eeq
Clearly, for the $W$ boson, $ m^2 = 2 \pi \, g^2 v^2$, whereas 
for the
 $Z$ boson, $m^2 = 2 \pi \, (g^2 + g'^2)v^2$,  with $g$ and $g'$ being
the weak and hyper-charge coupling constants respectively.

As for the fermions, six dimensions (unlike five) 
admit Weyl fermions, and we just promote the SM fermions to their 
higher-dimensional selves. 
Concentrating on the positive chirality spinor $\Psi_+$, 
the Dirac Lagrangian, in terms 
of the sechsbeins $E^{M}_{a}$ and spin connection $w_{M}^{bc}$,  is given by
\begin{equation}
{\cal L}_{\rm Dirac} = i \, \bar{\Psi}_{+} \, \Gamma^{a} \, E^{M}_{a} \, 
    \left( \partial_M + w_{M}^{bc}[\Gamma_b,\Gamma_c] \right) \, \Psi_{+} \ .
\end{equation}
Using a representation for the gamma matrices $\Gamma_b$ as in
Ref.\cite{Arun:2015kva}, the wavefunction $\Psi_{+}$ can be expressed as
\beq
\Psi_{+} = \frac{1}{\sqrt{R_y r_z}} \, \sum_{n,p}
     \left[{\cal F}^{n,p}_{+l}(x_4,x_5) \, \psi^{n,p}_{l}(x_{\mu}) \otimes \ \sup
       + {\cal F}^{n,p}_{+r}(x_4,x_5) \, \psi^{n,p}_{r}(x_{\mu})\otimes \sdn \right] \ ,
\eeq
with
\beq
\sup \equiv (1 \quad 0)^T \ , \quad
\sdn \equiv (0 \quad 1)^T \ . 
\eeq
A similar expression arises for $\Psi_{-}$ as well.
The subscripts ($l,r$) refer to the (four-dimensional) chirality of the 
four-dimensional fields $\psi^{n,p}_{l,r}$. 
Effecting a separation of variables, the wavefunctions 
${\cal F}^{n,p}_{+l/r}(x_4,x_5)$ can be written as
\begin{equation}
{\cal F}_{l/r}^{n,p}(x_4,x_5) 
= \left[a(x_4)\right]^{-2} \, \left[b(x_5)\right]^{-5/2}
\ft^{n,p}_{l/r}(x_4)f_{l/r}^{p}(x_5)
\end{equation}
where
\begin{equation}
\label{fly}
\barr{rcl}
\ft^{n,p}_l(x_4)& = & \dis e^{c|x_4|/2}\, 
           \left[c_1 J_{\nu_p}(x_{np}e^{c(|x_4|-\pi)}) 
               + c_2 Y_{\nu_p}(x_{np}e^{c(|x_4|-\pi)})  \right]
\\[1ex]
\ft^{n,p}_r(x_4)& = & \dis e^{c|x_4|/2}\,
           \left[ c_3 J_{\nu_p}(x_{np}e^{c(|x_4|-\pi)}) 
                + c_4 Y_{\nu_p}(x_{np}e^{c(|x_4|-\pi)})  \right]
\\[1ex]
\nu_p & \equiv & \dis \sqrt{\frac{1}{4} + \frac{m_p^2 R_y^2}{c^2}}
                      = \frac{p \, \pi}{2 \, \Theta_k(\pi)}
\\
x_{np} & \equiv & \dis M_{np} \frac{R_y}{c} e^{c \pi} \ ,
\earr
\end{equation}
and
\beq
\barr{rcl}
\label{flfr}
f_l(x_5) & = & 
 
   \, \exp\left[i \kappa_{l}^{+} \Theta_k(x_5)\right] 
-  \frac{d_l^+} {d_l^-}
\, \exp\left[i \kappa_{l}^{-} \Theta_k(x_5)\right] 
\\[1ex]
f_r(x_5) & = & 
  \, \exp\left[i \kappa_{r}^{+} \Theta_k(x_5)\right] 
-
 \, \exp\left[i \kappa_{r}^{-} \Theta_k(x_5)\right] 

\\[1ex]
\Theta_k(x_5) & \equiv & \dis \tan^{-1} \, 
\left(\tanh \frac{k x_5}{2}\right) \ .
\earr
\eeq
 The constants $\kappa_{l/r}$ are solutions of quadratic 
equations, and are given by
\beq
\barr{rcl}
\kappa_r^{\pm} & = & \dis 
-1 \pm \sqrt{1+4\frac{m_p^2R_y^2}{c^2}} 
\\[2ex]
\kappa_l^{\pm} & = & \dis 
1 \pm \sqrt{1+4\frac{m_p^2R_y^2}{c^2}} \ .
\earr
\eeq
For the massless mode,
$f_{l}^p(x_5) = 1$ and $\ft_{l}^{n,p}(x_4) = 1$. 
The boundary conditions dictate that $ {\cal F}_{+l}^{(n_1,p_1)}(x_4,x_5)
= {\cal F}_{-r}^{(n_1,p_1)}(x_4,x_5) $ and $ {\cal
F}_{+r}^{(n_1,p_1)}(x_4,x_5) = {\cal F}_{-l}^{(n_1,p_1)}(x_4,x_5)$.

The Yukawa Lagrangian now sees only the brane-localized Higgs field,
and can be written as
\beq
{\cal L}_y =  \sum_{i,j} 
   Y_{ij} \int d^4x\int dx_4\int dx_5 \sqrt{-g_5}\, 
    \phi^{\dagger} \, D^i_{+}(x_M) \, S^j_{-}(x_M) \, \delta(x_4-\pi) 
   + H.c. \ ,
\eeq
where $D_{+}^i(x_M) \, \big(S_{-}^j(x_M) \big)$ are the 
six-dimensional fields with chirality $\pm$
and transforming  as doublets (singlets) under $SU(2)$.

In terms of the KK components, this can be re-expressed as
\[
\barr{rcl}
{\cal L}_y &= & \dis
v \, \sum_{i,j}  \,
 Y^{n_1,p_1,n_2,p_2}_{(+l,-r)ij} 
\, \int d^4x \, D^{(n_1,p_1),i}_{+l}(x_{\mu}) \, S^{(n_2,p_2),j}_{-r}(x_{\mu})
\\[2ex]
&+ & \dis 
v \,  \sum_{i,j} \,
 Y^{n_1,p_1,n_2,p_2}_{(+r,-l)ij} 
\, \int d^4x \, D^{(n_1,p_1),i}_{+r}(x_{\mu}) \, S^{(n_2,p_2),j}_{-l}(x_{\mu})
+ H.c.
\earr
\]
where the effective four-dimensional Yukawa couplings are 
given by
\[
\barr{rcl}
\dis Y^{n_1,p_1,n_2,p_2}_{(+l,-r)ij} 
  & = & \dis Y_{ij} \, a^4(\pi) \, \int dx_5 \, \left[b(x_5)\right]^{9/2} \,
         {\cal F}_{+l}^{(n_1,p_1)}(\pi,x_5) \, 
         {\cal F}_{-r}^{(n_2,p_2)}(\pi,x_5)
\\[2ex]
\dis Y^{n_1,p_1,n_2,p_2}_{(+r,-l)ij} & = & \dis
    Y_{i,j} \, a^4(\pi) \, \int dx_5\, \left[b(x_5)\right]^{9/2} \,
         {\cal F}_{+r}^{(n_1,p_1)}(\pi,x_5) \,
         {\cal F}_{-l}^{(n_2,p_2)}(\pi,x_5) \ .
\earr
\]

Note that fermion mixing is, now, not restricted to just the
usual flavour (Cabibbo) mixing, but is generalized to incorporate
mixing between different KK excitations as well, both flavour-diagonal
and non-diagonal. This is but a consequence of the brane-localization of the 
Higgs field, which breaks KK number conservation. Concentrating 
on the inter-level mixing, while keeping the CKM mixing in abeyance for now, 
clearly the former is important primarily for the heaviest flavour, viz. the 
top-quark. The boundary
conditions ensures that the zero mode is chiral while leaving
the higher modes to be
vector like.  The mass matrix, in the weak/KK
eigenbasis 
 $Q_l=\Big(D_{+l}^{0,0},D_{+l}^{1,1},S_{-l}^{1,1} \Big)$
and $Q_r = \Big(S_{-r}^{0,0},D_{+r}^{1,1},S_{-r}^{1,1} \Big)$, reads
\[
{\cal M}_{top} = \begin{pmatrix}
           Y^{0,0,0,0} v  & 0 & 
	   Y^{0,0,1,1}_{(+l,-r)}v \\
           Y^{1,1,0,0}_{(+l,-r)}v
	   & M_{D(1,1)} & Y^{1,1,1,1}_{(+l,-r)}v \\
           0 &  Y^{1,1,1,1}_{(-l,+r)}v & M_{S(1,1)}
               \end{pmatrix}
\]
where $M_{D(1,1)}$ and $M_{S(1,1)}$ are the tree level KK masses (in
the absence of level-mixing) for the corresponding doublet and singlet
fields.  We have, obviously, truncated the mass matrix to the lightest
nontrivial sector, so as to illustrate the salient points without
unduly increasing the complexity. The physical masses are, of course,
given by the eigenvalues of ${\cal M}_{top}^\dagger {\cal
  M}_{top}$. Since the doublet and singlet masses are related by a
chiral rotation, $M_{D(1,1)} = - M_{S(1,1)} = M_{(1,1)}$, as
calculated in Paper I. On the other hand, the very structure of the
${\cal F}$'s ensure that, for a given fermion, the inter-level Yukawa
couplings are, generically, much smaller than the same-level
ones\footnote{Note that the inter-level couplings would have vanished
  if the Higgs field could percolate freely into the bulk and are but
  a consequence of the loss of KK-number conservation brought about by
  the brane localization.}.  In other words, $Y^{0,0,0,0} \approx
Y^{1,1,1,1}_{(+l,-r)}\approx Y^{1,1,1,1}_{(-l,+r)} \gg
Y^{0,0,1,1}_{(+l,-r)} = Y^{1,1,0,0}_{(+l,-r)}$, with the last
equality being an exact one.  This makes the diagonalization of the
matrix easier and, to the first order, similar to the Universal
Extra Dimension scenarios, with the caveat that, in warped space,
the Yukawa coupling constants are not all the same.  Though the
coupling increases for higher $p$ states, this is overshadowed by the
increase in the tree level KK mass. And hence we could truncate the
mass spectrum to $n=1,p=1$ level.  On diagonalizing the above matrix
numerically we get $M_{D(1,1)}^d = - M_{S(1,1)}^d \approx
\sqrt{M_{1,1}^2+\Big(\frac{Y^{1,1,1,1}_{(+l,-r)}+
    Y^{1,1,1,1}_{(-l,+r)}}{2}v \Big)^2}$.  For
the rest of the fermions, we could safely assume $M_{D(n,p)}^d = -
  M_{S(n,p)}^d \approx M_{n,p}$.

\section{ Higgs }
\label{sec:higgs}
A generic 3-brane localized Higgs profile leads to equations of motion
for the gauge bosons that do not admit simple closed form solutions,
and this is what prompted the particular choice\footnote{Note that the 
factor of $r_z^{-1/2}$ is only a overall normalization and is not reflective 
of the natural scale of the five-dimensional theory, which would
be seen to be $R_y^{-1}$.} of $\phi_{\rm cl} = v
/ \sqrt{r_z b(x_5)}$, in the previous section.  This seemingly ad hoc
  ansatz is actually a $x_5$-dependent solution\cite{Arun:2015kva} of
  the equation of motion for a potential of the form
\beq 
V(\phi) = \frac{k^2}{R_y^2}\left[ \frac{5 \; {\rm sech}^2 k
    \pi}{24 \,(v/\sqrt{r_z})^4} \, \phi^6 - \, \frac{7}{8} \, \phi^2
  \ \right] .
\label{effpotfirst}
\eeq
Note that this potential (proportional to that in
Ref.\cite{Arun:2015kva}) is truly of the aforementioned cutoff scale.

Perturbing the scalar field about its classical value, 
viz. $\phi(x_\mu, x_5) = \phi_{\rm cl}(x_5) + \hat{\phi}(x_\mu, x_5)$, 
we have, for the equation of motion
\[
\frac{1}{R_y^2} \, \partial_5(b^4\partial_5\hat\phi) 
+ \frac{b^2}{a_\pi^2}\partial_\mu\partial^\mu \hat\phi 
= \frac{k^2}{R_y^2} \, \Big(\frac{25 {\rm sech}^2 k \pi}{4} b^2 
           - \frac{7 }{4}b^4 \Big) \, \hat\phi \ .
\]
Re-parameterizing 
\beq
\hat\phi = \frac{1}{\sqrt{r_z}} \, h_p(x_\mu) \, \chi^{(h)}_p(x_5) \ ,
\eeq
we have
\beq
\barr{rcl}
\dis \partial_\mu\partial^\mu h_p & = & \dis m_p^2 h_p
\\[2ex]
\dis \partial_5 \left[b^4\partial_5\chi^{(p)}_p\right]
 & = & \dis - k^2 \, \left[\gamma_p b^2 + \frac{7}{4} b^4\right]\,
        \chi^{(h)}_p \ ,
\earr
\label{higgseqm}
\eeq
where 
\beq
m_p^2 = \Big( \frac{25}{4}{\rm sech}^2 k \pi+\gamma_p \Big)\frac{k^2 a_\pi^2}{R_y^2} \ .
\label{higgsmass}
\eeq 
Note that the nominal vacuum expectation value $v$ does not enter the
expression for the masses, but the cutoff $R_y^{-1}$ squarely does so;
and that the masses (as also $v$) are of the order of the
cutoff\footnote{This would not have been the case had we not effected
  the aforementioned scaling of the potential and the $x_5$ derivative
  term.}. A couple of
subtleties need to be considered, though. For one, the last term in
Eq.\ref{higgseqm}, namely $ \frac{7}{4} k^2 b^4$, could be considered
a negative ``bulk mass'' term. Furthermore, note that $\phi_{\rm cl}$
lives entirely on one side of the nominal vev $v$. Thus, despite the
positive contributions to the energy engendered by the nontrivial
$x_5$-dependence, there is a danger of the theory admitting tachyonic
modes (at least for some range of $k$), thereby invalidating the
formulation. We shall shortly return to this.

In the regime where all $m_p^2$ are non-negative, it is natural to 
identify the lowest state (corresponding to $\gamma_0$) with the recently
discovered Higgs boson, yielding
\beq
m_h = m_0 = \sqrt{\frac{25}{4}{\rm sech}^2 k \pi+\gamma_0 } \, 
    \frac{k}{R_y} \, a_\pi \ .
\label{higgsmasssqrt}
\eeq
Parameterizing the vev $v$ as
\beq
    v = \frac{\lambda_v}{\sqrt{2 \pi}} \, R_y^{-1} \, = \frac{\lambda_v}{\sqrt{2 \pi}} \, \aleph^{-1} \, r_z^{-1}\ , 
\label{lambda_v}
\eeq
where $\lambda_v \lapp 1$, we have
\beq
\lambda_v =  \sqrt{2\pi \, \Big(\frac{25}{4}{\rm sech}^2 k \pi+\gamma_0 \Big)}
    \; \frac{k}{g} \; \frac{M_w}{m_h} \ ,
\label{lambdaconstraint}
\eeq 
a relation that would prove to be useful in 
identifying the ``right'' part of the parameter space.
It should be remembered, though, that this result is only an
indicative one and can receive large corrections as we shall
see later.

\subsection{The Higgs Spectrum}
The solution to the equation of motion (Eq.\ref{higgseqm}) is given by
\beq
\barr{rcl}
\chi_{p}(x_5) & = & \dis 
   {\rm sech}^2(k z) \, \left[ 
     \cot\theta_p P_{\nu^{(h)}_p}^{3/2}(\tanh k x_5)
     + Q_{\nu^{(h)}_p}^{3/2}(\tanh k x_5)   \right]
\\[1ex]
\nu^{(h)}_p & \equiv& \dis \frac{-1}{2} + 
                    \frac{1}{2}\sqrt{9+4 \gamma_p \cosh^2(k\pi)} \ .
\label{higgssolution}
\earr
\eeq
Since the solutions have to be even 
functions of $x_5$, we have $\chi_p'(x_5 = 0) = 0$. Using
the identities
\[
\barr{rcl}
\dis \left( \frac{d P_N^M (x)}{d x} \right)_{x = 0}
  & = & \dis 
\frac{2^{M + 1}}{\sqrt{\pi}} \,
    \sin\left( \frac{\pi \, (N + M)}{2} \right) \,
    \frac{\Gamma( 1 + (N + M) / 2)} {\Gamma( (N - M + 1) / 2)}
\\[3ex]
\dis \left( \frac{d Q_N^M (x)}{d x} \right)_{x = 0}
  & = & \dis 
2^{M} \, \sqrt{\pi} \,
    \cos\left( \frac{\pi \, (N + M)}{2} \right) \,
    \frac{\Gamma( 1 + (N + M) / 2)} {\Gamma( (N - M + 1) / 2)}
\earr
\]
we are led to
\begin{equation}
\cot \theta_p = \frac{-\pi}{2} \, \cot \frac{\pi \, (\nu^{(h)}_p + 3/2)}{2} \ .
\label{cottheta}
\end{equation}
Since we are interested only in the small-$k$ branch, the Legendre 
functions are well-behaved in the entire domain and the use 
of the Neumann boundary conditions is straightforward, giving rise to
\begin{equation}
\barr{rcl}
0 & = & \dis 
(1-2\nu^{(h)}_p) \, \left[ \cot\theta_p \, P_{\nu^{(h)}_p+1}^{3/2}(\tau_\pi)
+ Q_{\nu^{(h)}_p+1}^{3/2}(\tau_\pi) \right]
 \\
&+& \dis 
2 \, (\nu^{(h)}_p-1) \, \tau_\pi \, 
\left[ \cot\theta_p \,P_{\nu^{(h)}_p}^{3/2}(\tau_\pi)
+ Q_{\nu^{(h)}_p}^{3/2}(\tau_\pi)\right]  \ ,
\earr
\end{equation}
where $\tau_\pi \equiv \tanh( k \, \pi)$. This equation has to be
solved numerically to obtain the discrete set of values allowed to $\nu^{(h)}_p$
and, hence, $\gamma_p$. 

Before we attempt this, it is amusing to note that a negative value
for $\gamma_0 $ would turn $\nu^{(h)}_0$ complex. Note though that
$Im\Big(P_{\nu^{(h)}_0}^{3/2}(\tanh(k
x_5))\Big)=Re\Big(Q_{\nu^{(h)}_0}^{3/2}(\tanh(k x_5))\Big)=0$, if
$Re\Big(\nu_0^{(h)}\Big) = - 1/2$, as is the case here.
In other words, the boundary conditions demand that, in such cases, 
$\cot \theta_p$ must be a pure imaginary number, as indeed is the case
(see Eq.\ref{cottheta})
The phase of the corresponding wavefunction would, thus, be independent 
of $x_5$. 

\begin{figure}[!h]
\vspace*{-50pt}
\centerline{
\epsfxsize=8cm\epsfbox{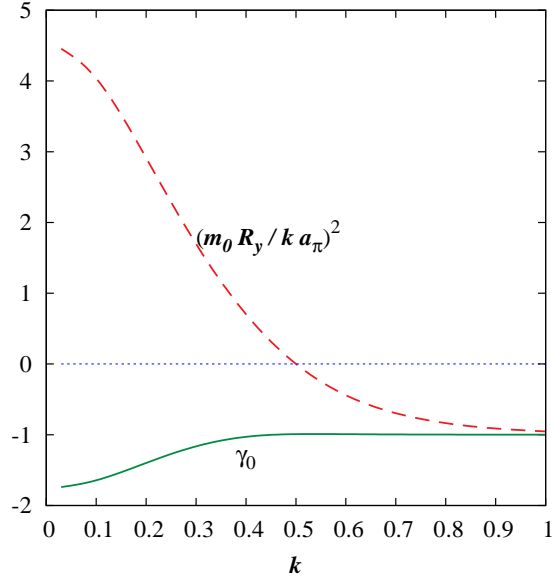}
}
\vspace*{-70pt}
\caption{\em The dependence of the lowest lying Higgs mass on 
$k$ starting from $\phi_{\rm cl} = v / \sqrt{r_z \, b(x_5)}$.} 
\label{fig:gamma_p}
\end{figure}

In Fig.\ref{fig:gamma_p}, we display the result for $\gamma_0$ as function 
of $k$. Also shown, for ready reference, is the dependence of the lowest 
mass $m_0$. As the figure clearly shows, the formulation allows for only 
$k \lapp 0.5$. On the other hand, a perusal of Table~\ref{tablehiggsmass}
(where the ratio $\aleph$ has been chosen to ensure that 
$m_0$ is consistent with the measured value) 
shows that requiring $\lambda_v \lapp 1$ (as argued for earlier)
would constrain us to $k \gapp 0.4$. This, then, seems to put 
strong constraints on the parameter space. It should be appreciated 
that the fast growth of the excited state masses with $k$ is but a 
consequence of the fact that, for such cases, the cancellation between 
the two pieces in the expression for $m_0$ is quite extensive, while this 
is not the case for the KK states. More interesting is the fact that 
$k \approx 0.45$ leads to a second scalar state mass of $\sim 700$--$800\gev$ 
as is indicated in the recent LHC results. 
As is obvious, the KK-excitation does not acquire a vev, 
and, hence, has drastically reduced partial width in to a $WW$ or $ZZ$ pair. 
On the contrary, its coupling with the top-quark (and its KK-cousins) 
remain unsuppressed, thereby leading to a much larger branching fraction 
into a $\gamma\gamma$ state. Consequently, it is an obvious candidate 
to explain the observed excess\cite{atlas_750, cms_750}. 
However, it should be 
realized that there is no conclusive evidence yet for such a resonance, and even less for its angular momentum.
\begin{table}[!h]
\begin{tabular}{*4{c}}
$
\begin{array}{|c|c|c|c|}
\multicolumn{4}{c}{\underline{k = 0.3, \; \alpha  = 49.0, \; w  =  2.69\times 10^{-14} }}
\\[1ex]
\hline
(p) & \gamma_{p} & m_{p} (\tev) & \\ 
\hline 
(0)&  -1.16419   & 0.121  & \lambda_v = 1.99, \beta =0 \\ 
\hline 
(1) &  4.932   & 0.260 & \\ 
\hline 
(2) & 24.742  &  0.489 & \\ 
\hline 
\end{array}
$
\hspace{1ex}
$
\begin{array}{|c|c|c|c|}
\multicolumn{4}{c}{\underline{k = 0.4, \; \alpha  = 46.5, \; w  =  4.33 \times 10^{-14} }}
\\[1ex]
\hline
(p) & \gamma_{p} & m_{p} (\tev) & \\ 
\hline 
(0)&  -1.029   & 0.120 &  \lambda_v = 1.71,\beta =0  \\ 
\hline 
(1) &  1.59 &   0.262 & \\ 
\hline 
(2) & 9.536 & 0.484 & \\ 
\hline 
\end{array}
$
\\[2ex]
$
\begin{array}{|c|c|c|c|}
\multicolumn{4}{c}{\underline{k = 0.5, \; \alpha  = 46 , \; w  =  3.2 \times 
10^{-13}  }}
\\[1ex]
\hline
(p) & \gamma_{p}& m_{p} (\tev) & \\ 
\hline 
(0)&  -0.9914   & 0.118 & \lambda_v = 0.323, \beta=-7 \times 10^{-6}\\ 
\hline 
(1) &  0.556 &   1.19 & \\ 
\hline 
(2) & 3.997 & 2.13 & \\ 
\hline 
\end{array}
$
\end{tabular}
\caption{\em Sample spectra for the small $k$ case 
for a particular bulk curvature ($\epsilon = 0.1$).}
\label{tablehiggsmass}
\end{table}

\subsection{Corrections to the Higgs potential and modifications to the 
spectrum}
That aesthetic considerations (as also phenomenological imperatives 
as we shall see soon) drive us towards a precipice in the parameter 
space (as exemplified by a possible tachyonic mode) behoves us to 
pause and reconsider. Is this a generic feature of the scenario or 
is it specific to the form of the potential that we have chosen? 
Even if Eq.\ref{effpotfirst} indeed represented the tree-level 
potential, it would, at the least, be subject to quantum corrections. 
In fact, given that we are dealing with a non-renormalizable theory 
(with a well-specified cutoff $R_y^{-1}$), we could as well consider 
higher-dimensional terms even in the tree-order Lagrangian. We will,
for the sake of simplicity, limit ourselves to polynomial terms. 

Even with a generic polynomial modification to the potential, an exact 
closed-form solution to the equation of motion is not straightforward. 
Furthermore, the specific form of $\phi_{\rm cl}$ was chosen to facilitate 
the solution of the gauge boson wavefunctions with the boundary-localized 
symmetry breaking term. To this end, we would like to preserve this 
feature to the best of our abilities and, thus, contemplate only a 
monomial\footnote{While a polynomial change is but a straightforward 
generalization of the analysis present here, it adds little to the 
qualitative features.} perturbation to $\phi_{\rm cl}$ of the form
\beq
\phi_{\rm cl}^{\rm new} = \frac{v}{\sqrt{r_z \, b(z)}} \, \left[ 1+ \beta_n b^n(z) \right]
\label{newphicl}
\eeq
where $n$ is an as yet undetermined power and $\beta_n$ is a small parameter. 
It is easy to see that the change above can be wrought about with a 
potential 
\beq
V_{new} (\phi) = V(\phi) + \delta V
\eeq
where 
\beq
\delta V = -\frac{k^2}{R_y^2}\,  \frac{v}{2 r_z^2} \, \beta_n 
    \left[ \frac{n^2+3n}{n-1} \left(\frac{r_z\, \phi^2}{v^2} \right)^{1-n}
     + \frac{n^2+2n+5}{3-n}{\rm sech}^2 (k\pi) \,
               \left(\frac{r_z\, \phi^2}{v^2} \right)^{3-n} \right] 
     + {\cal O}(\beta_n^2) \ .
 \label{pot_pert}
\eeq
Considerable simplification occurs for $n= -3$ (a choice that we embrace for the rest of the paper), whence the potential simplifies to
\[
\delta V = \frac{-4 \, k^2}{3 \, R_y^2}\,  \frac{v}{2 r_z^2} \, \beta \, 
     {\rm sech}^2 (k\pi) \,
               \left(\frac{r_z\, \phi^2}{v^2} \right)^{6} 
\]
where $\beta \equiv \beta_{-3} \leq 0$ so as to ensure a potential bounded 
from below. 

Perturbing around $\phi_{\rm cl}^{\rm new}$, the new equation of
motion is found to be
\beq
\frac{1}{R_y^2} \, \partial_5(b^4\partial_5\hat\phi) 
+ \frac{b^2}{a_\pi^2}\partial_\mu\partial^\mu \hat\phi 
 =  \frac{k^2}{R_y^2} \, \left[ \frac{25 {\rm sech}^2 k \pi}{4} b^2 
           - \frac{7 }{4}b^4 + \frac{\beta}{b(z)} 
              \left(25-\frac{28 \aleph}{\lambda_v}\right)
                   {\rm sech}^2 (k\pi) \right] \, \hat\phi \ .
\label{eqmotionphinew}
\eeq 
Treating the last term above as a perturbation, the lowest eigenvalue 
is shifted to
\[
m_{0, {\rm new}}^2 = \frac{k^2 a_\pi^2}{R_y^2} \, 
\left[ \Big( \frac{25}{4}{\rm sech}^2 k \pi+\gamma_p \Big)+
\beta\, \left(25-\frac{28}{v r_z}\right) X_{k} \right]
\]
where $X_k$ is the matrix element of the perturbation Hamiltonian. The
consequent shift in the wavefunction $\chi_0(x_5)$ can be calculated
analogously. For $k=0.5 \, (0.6)$ we have $X_k = 0.51 \, (0.3)$.
Clearly for $\lambda_v < 1$ (as it should be), a negative $\beta$
raises the Higgs mass considerably, thereby allowing for a wider range
of $k$ without risking tachyonic modes.  In the new description
Eq.\ref{lambdaconstraint} will get modified and, as
Fig.~\ref{fig:lam_bet}
shows, a rather wide range of $\lambda_v$ becomes allowed once even
small perturbations are switched on.
\begin{figure}[!h]
\vspace*{0pt}
\centerline{
\epsfxsize=8cm\epsfysize=9cm\epsfbox{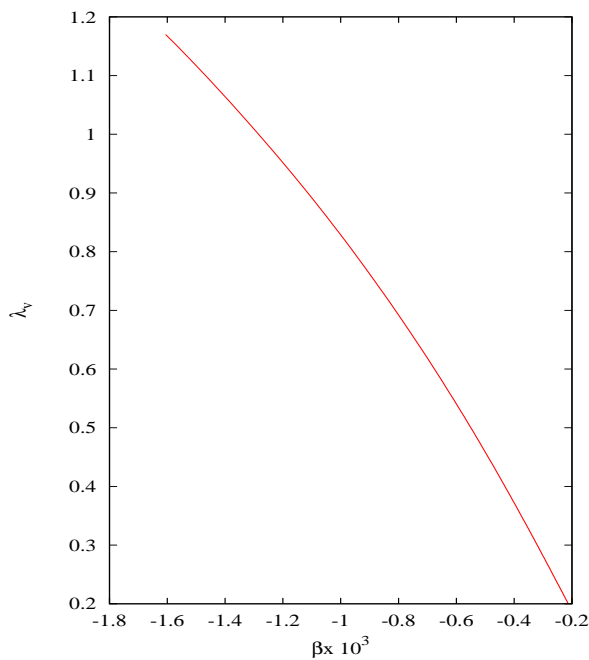}
}
\vspace*{-10pt}
\caption{\em The relation between the parameters $\lambda_v$ and the
coefficient $\beta_{-3}$ in the monomial 
scalar field perturbation ($\phi^{12}$) that would lead to a 
light Higgs mass of $125\gev$ for $k=0.6$ and $50 < \aleph < 53$.}
\label{fig:lam_bet}
\end{figure}

\subsection{An alternative scheme}
Appealing to corrections to the Higgs potential is not the only way out 
of the tachyonic imbroglio. We briefly consider, here, an alternative. 
Recall that the very establishment of the nested warping structure required 
a $x_5$-dependent tension on the 4-brane at $x_4 = \pi$ given by
\[
    V_2(x_5) = - 8 \, M_6^2 \, \sqrt{\frac{-\Lambda_6}{10}} \, 
                {\rm sech}(k \, x_5)
            = - 8 \, \epsilon \, M_6^5 \, {\rm sech}(k \, x_5) \ .
\]
The particular form for $V_2(x_5)$ could have originated from a 
variety of mechanisms including a $x_5$-dependent vacuum structure 
in a scalar field theory~\cite{Choudhury:2006nj}. As can be appreciated,
this is intimately connected to the very process of compactification 
in this theory. It is, thus, conceivable that such a dynamical system 
(whatever be the exact mechanism) could couple to the scalar $\phi$ 
as well. Thus we may posit a scalar field Lagrangian of the form
\beq
\widehat{\cal L}_\phi = \delta(x_4-\pi) \sqrt{-g_5}
    \left[ \frac{- \gamma \, V_2(x_5)}{M_6^5} \,
           \left\{ g^{\mu \nu}D_{\mu} \phi^{\dagger}(x^{\bar M})
                    D_{\nu} \phi(x^{\bar M}) + R_y^{-2} \, 
                        \left|D_{5} \phi(x^{\bar M})\right|^2
                                 \right\}
+ \widehat V(\phi) \right]
\label{alt_lagrangian}
\eeq
where $\gamma$ is a dimensionless positive constant. Choosing a 
standard form for $\widehat V(\phi)$, namely 
\[
\widehat V(\phi) = -\mu^2 \, \phi^\dagger \phi 
                 + \frac{\widehat\lambda}{2} \,(\phi^\dagger \phi)^2 \ ,
\]
would lead to a flat (i.e., $x_5$-independent) classical configuration
viz. $\widehat \phi_{\rm cl} = \widehat v = \sqrt{-\mu^2 / \widehat\lambda}$. 
The corresponding localized mass term for the gauge field 
is exactly what we get for the unperturbed potential $V(\phi)$ discussed 
earlier. On the other hand, with the scalar field $\phi$ now settling to its 
global minimum, and with the $x_5$-dependence of the fluctuation $\hat \phi$ 
(around $\widehat \phi_{\rm cl} = \widehat v$) only adding to the energy, 
no tachyonic modes exist any longer. This allows us to use a much wider range 
of $k$. 

It should be realized that these results are not tied to the exact 
form of $\widehat {\cal L}_\phi$, but would be replicated, to a great 
extent, for many other choices (for both the 
kinetic and the potential terms). This fact, as well as the results drawn, 
in the 
preceding section, from perturbing the potential leads us to the inescapable
conclusion that Eq.\ref{lambdaconstraint} is not an exact relation
but only an indicative one. This does not come as surprise, 
for once radiative corrections (whether in the 
full theory, or in the effective four-dimensional version) to the Higgs 
potential are taken into account, tree-level relationships would indeed
change (as happens, for example in the well-known case of the 
minimal supersymmetric standard model).

\section{Effective Lagrangian}
\label{sec:l_eff}
To examine the low-energy consequences of this model, and 
especially those of the electroweak symmetry breaking mechanism,
it is useful to construct an effective Lagrangian, which we 
do now. As we have already seen, the
resolution of the hierarchy problem with 
a localized (whether on a 3-brane or a 4-brane, as done here)
 Higgs boson
introduces nontrivial alterations to the boundary
conditions of the gauge bosons. In particular, such a
localized energy density deforms not only the mass spectrum, but also the wavefunctions of the
KK-modes.
However, since the symmetry-breaking mass is much smaller than the 
KK-masses, it is safe to consider the changes in the lowest (zero-) mode 
alone, while neglecting those to the others. Moreover, as we shall
soon see, the effect of such changes in the KK-mode wavefunctions 
on low-energy observables are further suppressed.

As we have already learnt, in the absence of the Higgs vev, the 
zero-mode wavefunctions for the gauge boson, viz. $\chi_0(x_5)$ and 
$\eta_{0,0}(x_4)$, are both flat. On inclusion of the 4-brane 
localized vev $\phi_{\rm cl}(x_5) = v / \sqrt{r_z \, b(x_5)}$, the latter 
changes to $(V \equiv W^\pm / Z)$
\beq
 \eta^V_{0,0} \rightarrow \eta^{Vr}_{0,0} \approx \frac{1}{\sqrt{\pi}}\, 
             \left[ 1 + \frac{M_V^2 \, \rho^2}{4} \, 
                \left( e^{2c(x_4-\pi)}-1-2cx_4e^{2c(x_4-\pi)}+2c\pi \right)\right] \ .
\label{eq:gaugeprofilechange}
\eeq
where
\beq
\rho = \frac{R_y}{c} \, e^{c \pi }
\label{eq:rho}
\eeq
and we are working 
under the approximation that $m_{0,0} \, R_y / c \ll 1$.

Before we use Eq.\ref{eq:gaugeprofilechange} to calculate any
observables, we should also consider changes wrought by the inclusion
of the perturbation of Eq.\ref{pot_pert} that would have led to a change in the classical
configuration encapsulated in Eq.\ref{newphicl}.  The effect of
this change in the boundary-localized energy density for the massive
gauge bosons can be calculated easily in perturbation theory, and, to
the first order in $\beta$, the zero-mode wavefunction changes to
\[
\eta^{Vr}_{0,0}(x_4) \to 
   \eta^{Vr}_{0,0}(x_4) + 
\left[ \frac{2 \,\beta_n \, v^2 \, a_{\pi}^2}{\pi \, (m_z^2-M_{(1,0)}^2)}
                    \int_{-\pi}^{\pi} dx_5 \, b^{n+1}(x_5) \, 
                       \left[\chi_0(x_5) \right]^2 \, \right]
                       \eta^{Vr}_{1,0}(x_4) + \cdots
\]
where the ellipsis denote the sub-dominant terms. 
With the integral being $\lapp {\cal O}(1)$, the additional suppression 
of $v^2 / M_{(1,0)}^2$ renders this correction too small to be of any interest,
and we shall neglect it altogether henceforth.

Reverting to Eq.\ref{eq:gaugeprofilechange}, such distortions
  manifest themselves, on integrating out the extra dimensions, as
  wavefunction renormalizations. On
canonically normalizing the kinetic term in the Lagrangian, this brings forth tree level
modification in the gauge mass term as also any
gauge interaction terms.
The relevant part of the renormalized Lagrangian, for the renormalized zero mode $V^r$, can be
written in terms of the self energy corrections $\Pi_{VV}(q^2)$
as
\[
{\cal L}_{V} =  - \frac{1}{4} {\cal Z} F^{r \mu \nu}F_{\mu \nu}^r
         - \frac{1}{2}\left[ M_V^2 + \Pi_{VV}(0)\right] \, V^{r2}, 
\]
where $M_V$ arises from the  Higgs vev and
\[
\barr{rcl}
\Pi_{VV}& = & \dis \left\{ \int_0^\pi dx_4 \, a^2(x_4) \, 
                          \left[\partial_4 \eta^{Vr}_{0,0}(x_4)\right]^2
              \right\} \;  
              \left\{ \int_0^\pi dx_4 
                          \left[\eta^{V}_{0,0}(x_4)\right]^2
              \right\}^{-1}
\\[2ex]
{\cal Z} & = & \dis 1 - \Pi'_{VV}(0) =  
                   \left\{\int_{0}^{\pi}dx_4 \left[\eta^{Vr}_{0,0}(x_4)\right]^2 
                   \right\} \; 
                   \left\{\int_{0}^{\pi}dx_4 \left[\eta^V_{0,0}(x_4)\right]^2 
                   \right\}^{-1} \ .
\earr
\]
Here, $\Pi'_{VV} \equiv \partial_{q^2} \Pi_{VV}$.
Note that, since the wave function in the
$x_5$--direction 
remains constant and unchanged, there is no corresponding
contribution to $\Pi_{VV}$ or $\Pi'_{VV}$.

Post electroweak symmetry breaking, we are primarily interested in the 
lowest modes, and the relevant part of the 
mass matrix can be diagonalized by a transformation analogous 
to that in the SM, viz.
\[
Z_{\mu} = c_\theta W_{\mu}^3 - s_\theta B_\mu \ ,
\qquad
 A_\mu = s_\theta W_{\mu}^3 + c_\theta B_\mu \ , 
\qquad
c_\theta \equiv \frac{g}{\sqrt{g^2+g'^2}} \ .
\]
Here, we neglect the small mixing with
 the higher KK-levels, 
which constitutes an excellent approximation.
In the basis where the
mass matrix is diagonal, the relevant part of the Lagrangian could be written as
\beq
\label{effective}
\barr{rcl}
- {\cal L}_{eff} & = & \dis 
    \frac{{\cal Z}_\gamma}{4} \, F_{\mu \nu}F^{\mu \nu}
   + \frac{{\cal Z}_W}{2} W^{+}_{\mu \nu}W^{-\mu \nu}
   + \frac{{\cal Z}_Z}{4} Z_{\mu \nu}Z^{\mu \nu} 
\\[1ex]
& +& \dis \left[M_w^2+ \Pi_{ww}(0)\right] W_\mu W^\mu 
  + \frac{1}{2} \, \left[M_z^2 +\Pi_{zz}(0)\right]\, Z_\mu Z^\mu
\\[2ex]
& \equiv & \dis \frac{1+A}{4} F_{\mu \nu}F^{\mu \nu}
    +  \frac{1+B}{2} W^{+}_{\mu \nu}W^{-\mu \nu}
    + \frac{1+C}{4} Z_{\mu \nu}Z^{\mu \nu}
    + \frac{G}{2}F_{\mu \nu}Z^{\mu\nu}
\\[2ex]
    & + & \dis (1+w) M_{W}^2 W_\mu W^\mu + \frac{1+z}{2} M_{z}^2 Z_\mu Z^\mu \ ,
\earr
\eeq
where we have deliberately introduced the parameters 
$A,B,C,G,w,z$ for future ease. 

Since the photon does not couple to the
Higgs, $\Pi_{\gamma\gamma} = 0$ and 
the corresponding renormalization factor ${\cal Z}_{\gamma} =
1$. For the $W$ and $Z$, we get instead
\[
\barr{rclcl}
 \Pi_{ww}(0) & = & \dis 
    \frac{1}{2} M_{w}^2 \Big(M_w \rho\Big)^2 \,
     \Big(\frac{1}{2 c \pi} -1 +  c \pi \Big) 
  & \equiv & \dis \frac{1}{2} M_{w}^2 \tilde{m}_w
\\[2ex]
 \Pi_{zz}(0) & = & \dis 
     \frac{1}{2} M_{z}^2 \Big(M_z \rho\Big)^2 \, 
    \Big(\frac{1}{2 c \pi} -1 +  c \pi \Big) 
  & \equiv & \dis \frac{1}{2} M_{z}^2 \tilde{m}_z
\earr
\]
which also implies that 
\[
\barr{rclclcl}
 {\cal Z}_w & = & \dis 1 - \Pi'_{ww} & = & \dis 
    1 - g^2\Pi'_{11} & \approx & \dis 1+ \tilde{m}_w
\\[2ex]
  {\cal Z}_z & = & \dis 1 - \Pi'_{zz} & = & \dis 
   1 - (g^2+g'^2)\Pi'_{33} & \approx & \dis 1+ \tilde{m}_z \ .
\earr
\]
With the rest of the gauge sector 
unchanged, the gauge-fermion interaction
can now be expressed in the standard form, viz.
\beq
 \label{fermiongaugeint}
{\cal L}_{int} = \left[
   g \sum_{i,j} V_{ij} \bar{\psi}_i\gamma^{\mu} P_{L}\psi_j  W^{+}_{\mu} + H.c \right]
 +  \frac{g}{c_\theta} \sum_i \bar{\psi}_i\gamma^{\mu}\Big( T_{3i} P_L - Q_i s_{\theta}^{2} \Big)\psi_i Z_{\mu}+ 
 e \sum_i \bar{\psi}_i\gamma^{\mu}Q_{i}\psi_i A_{\mu} \ ,
\eeq
with all modifications encoded in the aforementioned six parameters $A,B,C,G,w$ and $z$.
Note, though, that, on redefining
$W_{\mu}^a$, $B_\mu$ and Higgs fields, only three
of the six would remain independent and have been famously 
parametrized as $S,T,U$~\cite{Peskin:1991sw} (or, equivalently, 
$\epsilon_{1,2,3}$~\cite{Altarelli:1990zd}) through the relations
\beq
\label{oblique}
\barr{rcl}
\alpha_{\rm em} S & \equiv & \dis 4 s_{\theta}^2 c_{\theta}^2 
   \left[ A- C - \frac{c_{\theta}^2-s_{\theta}^2}{s_{\theta}c_{\theta}}G \right]
\\[1ex]
\alpha_{\rm em} T & \equiv & \dis w - z
\\[1ex]
\alpha_{\rm em} U & \equiv & \dis  4 s_{\theta}^2\,
   \left[ A- \frac{B}{s_{\theta}^2} +\frac{c_{\theta}^2}{s_{\theta}^2} C - 2 \frac{c_{\theta}}{s_{\theta}} G \right]
\earr
\eeq

\subsection{The Oblique Parameters}
\label{sec:obliqueparameter}
Having considered the general form, we now concentrate on the
particular case at hand, namely the {\em extra} corrections wrought by
the new physics over and above the SM contributions, with the latter
accruing only at the loop level. On the contrary, the additional
contributions here are two-fold. One set is
occasioned by the exchange of the KK-excitations and these we shall
come back to later. The other is occasioned by a change in the
wavefunctions of the SM particles and appear even at the
tree-level. Given this, we may as well neglect any loop-level effects
associated with the new physics.  This approximation immediately leads
to certain simplifications.  For example, consider $\Pi_{3Q}$, which,
in the SM, is generated only at the loop-level. Since ours is a
tree-level calculation of the new physics effect, no additional
$Z$--$\gamma$ mixing can be induced ($\delta\Pi_{3Q} = 0$). This, of
course, was evident from Eq.\ref{effective} as it implied $G=0$.

Renormalizing the fields through
\[
 A_{\mu} \rightarrow A_{\mu}^r = A_{\mu} \ , \qquad \quad
 Z_{\mu} \rightarrow Z_{\mu}^r=\sqrt{{\cal Z}_z} \, Z_{\mu} \ , \qquad\quad
 W_{\mu} \rightarrow W_{\mu}^r=\sqrt{{\cal Z}_w} \, W_{\mu} \ ,
\]
the gauge kinetic term can be expressed as
\beq
\barr{rcl}
{\cal L}_{eff} = - \frac{1}{4} F^r_{\mu \nu}F^{r\mu \nu}- \frac{1}{2} W^{r+}_{\mu \nu}W^{r-\mu \nu}- \frac{1}{4} Z^r_{\mu \nu}Z^{r\mu \nu} + 
M_w^{r2} W^{r2} + \frac{1}{2}M_w^{r2} Z^{r2} \ ,
\earr
\eeq
where the renormalized masses are given by
\[
\barr{rclcl}
M_w^{r2} & = & \dis M_{w}^2( 1+w-B) & = & \dis M_w^2 \, 
                \left[1-\frac{1}{2}\tilde{m}_w \right]
\\[1ex]
M_z^{r2}& = & \dis M_{z}^2( 1+z-C) & = & \dis M_z^2\, 
                \left[1-\frac{1}{2}\tilde{m}_z\right] \ .
\earr
\]
Similarly, the gauge fermion interaction is given by
\beq
\barr{rcl}
 {\cal L}_{int} & = & \dis 
 \left[\frac{g}{\sqrt{{\cal Z}_w} } \,
                       \sum_{ij} V_{ij} \bar{\psi}_i\gamma^{\mu} P_{L}\psi_j \, 
                           W^{r+}_{\mu} + H.c. \right]
\\[2ex]
               & + & \dis 
\frac{g}{c_\theta \, \sqrt{{\cal Z}_z} } 
                   \, \sum_i \bar{\psi}_i\gamma^{\mu} \,
                       \left( T_{3i} P_L - Q_i s_{\theta}^{2} \right) \, 
                          \psi_i  Z^{r}_{\mu}
               + e \sum_i \bar{\psi}_i\gamma^{\mu}Q_{i}\psi_i A^r_{\mu} \ .
\earr
\eeq
This immediately leads to expressions for the oblique parameters
\beq
\barr{rclcl}
\delta S & \approx & \dis - 4 \pi \frac{M_w^2 \rho^2 c \pi}{g^2}  & = & - 4 \pi \zeta
\\[2ex]
\delta T & \approx & \dis 
      \frac{-\pi}{2 \cos^2\theta_w} \,  \frac{M_w^2 \rho^2 c \pi}{g^2}  
          & =  & \dis \frac{-\pi}{2 \cos^2\theta_w} \zeta
\\[1.5ex]
\delta U & = & 0
\\[1.5ex]
\zeta & \equiv & \dis  \frac{ M_w^2 \rho^2 c \pi }{g^2} \ .
\earr
\eeq
A detailed fit to the data has been performed in \cite{Agashe:2014kda}, and we use their central values ( derived by fixing U=0, as is the case here and as is normal for most beyond-SM fits) of $S= 0.00 \pm 0.08$ and $T = 0.05 \pm 0.07$.

\subsection{$G_f$}
\label{sec:fourfermi}
In the most popular renditions of the SM fields leaking into a {\em
  flat} bulk (the so-called Universal Extra Dimension scenarios), the
existence of a $Z_2$ symmetry prevents the odd KK-modes of the gauge
bosons from coupling with the SM bilinear. Furthermore, the couplings
of the even-modes are progressively suppressed for the higher
modes. No such symmetry exists here, and all modes of the gauge-bosons
would couple with non vanishing strengths to the zero-mode fermion
bilinear.  In particular, the coupling of the $(1,0)$--mode is often
enhanced with respect to the SM coupling.  This immediately leads to a
change in the four-fermion operators.  For charged current processes
at low energies, this is parametrized by the very well measured
quantity $G_f$ which now reads
   
\beq
G_f = G_f^{\rm SM} \, \left[ 1 + \sum_{(n,p) \neq (0,0)} 
                             \left( \frac{g^{(n,p)} \, M_W}{g \, M_{W\,(n,p)}}
                             \right)^2 \right]
   \approx G_f^{\rm SM} \, \left[ 1 + 
                             \left( \frac{g^{(1,0)} \, M_W}{g \, M_{W\,(1,0)}}
                             \right)^2 \right] \ .
\eeq    
To appreciate the approximation above, it should be 
remembered that, for a given $p$, it is the coupling of 
the $n = 1$ mode, viz. $g^{(1,p)}$, that is the largest, while those 
for the higher $n$-modes are, typically, somewhat suppressed 
with respect to the SM coupling (see Table 1 of ref.\cite{Arun:2015kva}). 
Compounded by the fact that the higher modes are much heavier, 
it is clear that, within the $p = 0$ tower, the contribution of the 
$n = 1$ mode dominates. For $p \neq 1$ modes, all the couplings are 
significantly suppressed (even for $n = 0$) and the masses larger.

In other words, we have
\beq
G_f \approx G_{f}^{\rm SM} \, \left[
     1 + V \right] \ ,
 \qquad
V \equiv \frac{\zeta}{\pi \, c} \, \left( \frac{g^{(1,0)}}{x_{1,0}} \right)^2 \ .
\eeq
Experiments demand~\cite{Davoudiasl:1999tf} $V < 0.0013$ at 95\% C.L.
\begin{table}[!h]
\begin{center}
$
\begin{array}{|c|c|c|c|}
\multicolumn{4}{c}{\underline{k = 0.5, \; \alpha  = 48.367, \; w  =  7.081 \times 10^{-14} }}
\\[1ex]
\hline
(n,p) & m_{np} (\tev) & C_{np} & V \\
\hline 
(1,0)&   9.5    & 3.81  & 1.0 \times 10^{-3}\\ 
\hline 
(2,0) &  21.9 & 0.49  & 3.34 \times 10^{-6}\\ 
\hline 
(0,1)& 17.0 & 0.20  &  9.21 \times 10^{-7} \\ 
\hline 
(1,1) & 30.9  &  0.06 & 2.84 \times 10^{-8}\\ 
\hline 
\end{array}
$
\caption{\em Sample spectrum for the small $k$ case 
for a particular bulk curvature ($\epsilon = 0.1$) and with 
$\lambda_v = 1.5$.
$C_{np}$ is defined as the ratio of $g^{(n,p)}$ 
and $g$.}
\label{tab:small_k}
\end{center}
\end{table}

To a reasonable degree of accuracy, the coupling of the $W^{\pm(1,0)}$ 
to the fermion bilinears could be approximated as 
$g^{(1,0)} \sim 3.8 \times g$. Using this, we have 
$M_{W^{(1,0)}} \gapp 8.6 \tev$, a constraint that is a little weaker than 
that operative for the RS case. This was not unexpected, because 
the suppression of the gauge-excitation coupling (in relation the 
five-dimensional analogue), is only a small one, as evinced by the 
aforementioned approximation. What is more interesting is that, as 
Table.\ref{tab:small_k} shows, there exists a large parameter space
where this constraint is automatically satisfied. We will delineate 
this quantitatively in the next section.

\section{Confronting Electroweak Precision Measurements}
\label{sec:chisq}
Rather than drawing conclusions piecemeal from individual data (as we
have done in the preceding section), we now attempt to examine how
well the model agrees globally with all the precision measurements.
Ref.\cite{Csaki:2002gy} drew up expressions for 22 such observables in
terms of the their SM values, the oblique
parameters~\cite{Peskin:1991sw} $S,T,U$ and $V$ (the shift in $G_F$).
While the extra-dimensional contributions to the $U$-parameter are
vanishingly small, for $S$, $T$ and $V$, we use the expressions
derived in the preceding section.  Re-evaluating the SM expectations for
a $125 \gev$ Higgs\footnote{More up-to-date analyses, including two-loop
  results, are available~\cite{Baak:2012kk,
    Ciuchini:2013pca,Baak:2014ora}, but make little qualitative
  difference to our conclusions.},  we may now construct a
$\chi^2$-test for this model comparing the expressions with the
experimental results \cite{Agashe:2014kda}.

\begin{figure}[!h]
\vspace*{-20pt}
\centerline{
\epsfxsize=7cm\epsfysize=10cm\epsfbox{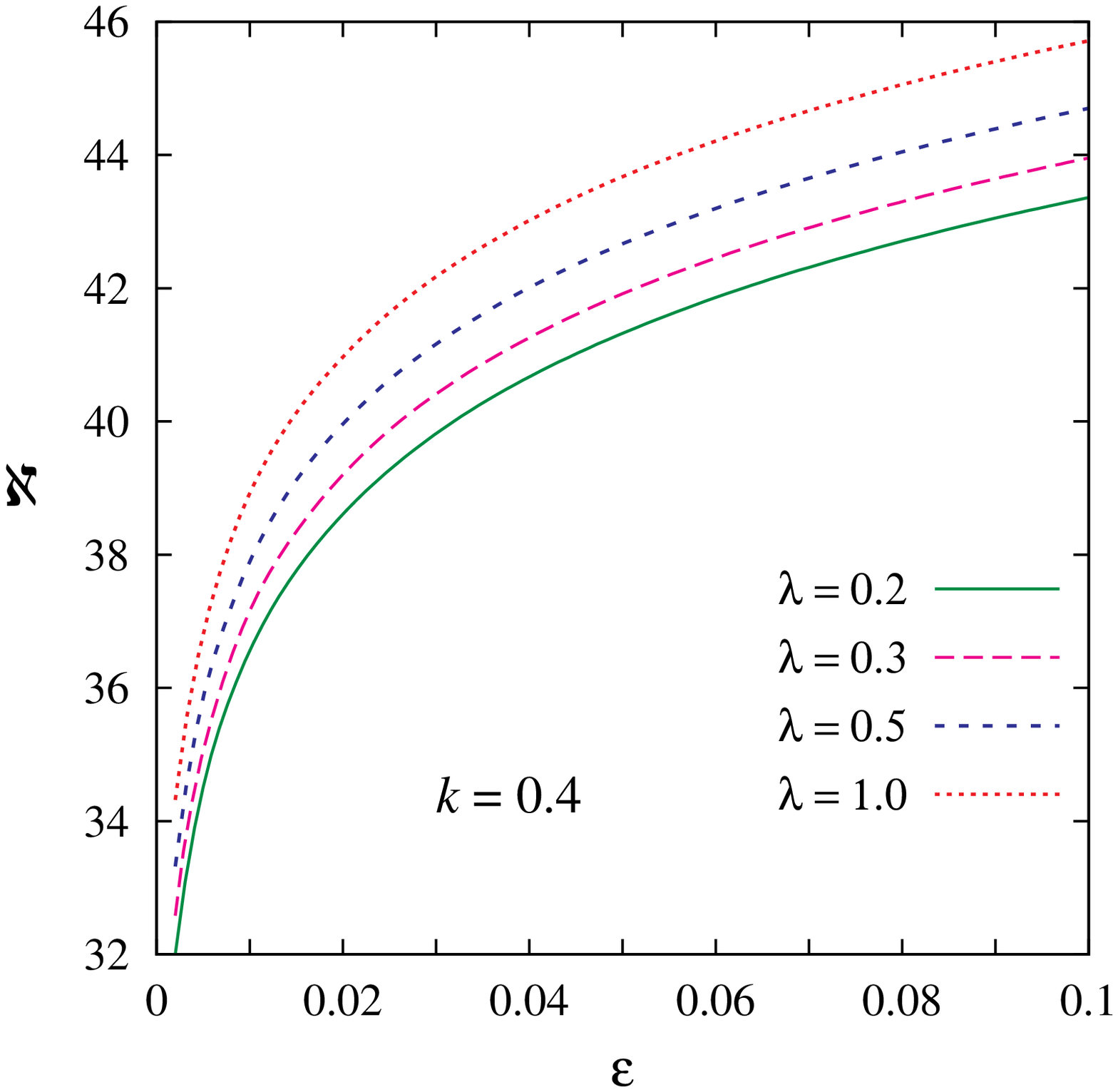}
\epsfxsize=7cm\epsfysize=10cm\epsfbox{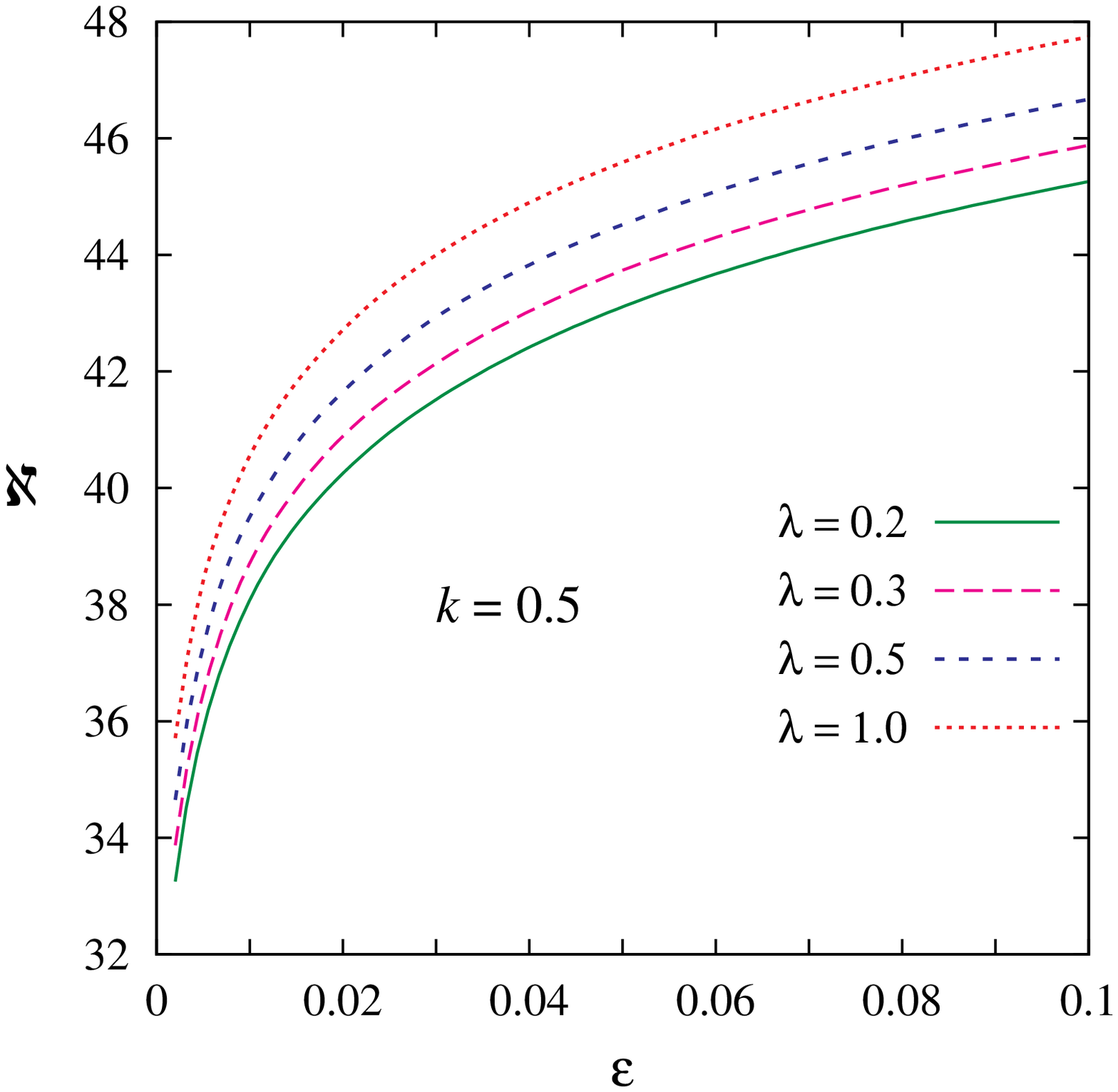}
}
\vspace*{-70pt}
\caption{\em The relation between the parameters $\epsilon$ and
  $\aleph$ that reproduces the electroweak breaking scale
  correctly. The left and right panels refer to $k = 0.4$ and $k =
  0.5$ respectively.}
\label{fig:epsi_aleph}
\end{figure}
While one could attempt a multidimensional analysis optimizing all the
parameters in the theory, it is much more instructive to examine the
dependence of the $\Delta\chi^2$ (the shift in the $\chi^2$ from the
SM value of $\approx 27.5$) on individual parameters. To this
  end, we must first identify the appropriate set of independent
  parameters, and the range that they may be allowed.  To start with,
  Eq.\ref{RS6_eqns} imposes two independent relations between $c,
  \aleph, k, \epsilon$ and the product $r_z M_6$.  Now, the
  applicability of a semi-classical treatment of the gravity sector
  requires that the curvature be sufficiently smaller than the
  fundamental scale $M_6$, or in other words, $\epsilon \lapp 0.1$.
  Similarly, the avoidance of a large hierarchy implies that the
  product $M_6 r_z$ be not too large. Since we are interested in the
  small $k$ regime ($k \lapp 1$), this immediately puts a lower bound
  on $\epsilon$. A complementary relation is provided by
  Eq.\ref{lambda_v}, and, once the electroweak scale is specified,
  the resultant relation between the parameters is determined as
  displayed in Fig.~\ref{fig:epsi_aleph}. As is apparent, the
  dependence on $k$ is minimal, owing to the fact that the function $k
  / \cosh(k \pi)$ is slowly varying in the region of interest.

In Fig.\ref{fig:epsi_chisq}, we present the corresponding shifts
  $\Delta\chi^2$. Understandably, the dependence on $k$, once again,
  is minimal. As Eq.\ref{lambda_v} shows, a smaller $\lambda_v$
  would imply a larger $R_y^{-1}$. This, in turn, has two
  consequences.  First, it implies larger masses for the
  KK-excitations of the gauge boson masses and, hence, a smaller
  change to $G_f$. Simultaneously, it results in smaller values for
  $\rho$ (see Eq.\ref{eq:rho}), and, hence, smaller values for both
  $\delta S$ and $\delta U$. Thus, it is easy to understand the
  dependence of $\Delta\chi^2$ on $\lambda_v$. It is interesting to
  note that even a very moderate hierarchy ($\lambda_v \lapp 0.3$)
  renders the model quite consistent with low-energy data, whereas
  $\lambda_v \sim 0.2$ makes it almost indistinguishable from the SM.


\begin{figure}[!h]
\vspace*{-20pt}
\centerline{
\epsfxsize=7cm\epsfysize=9cm\epsfbox{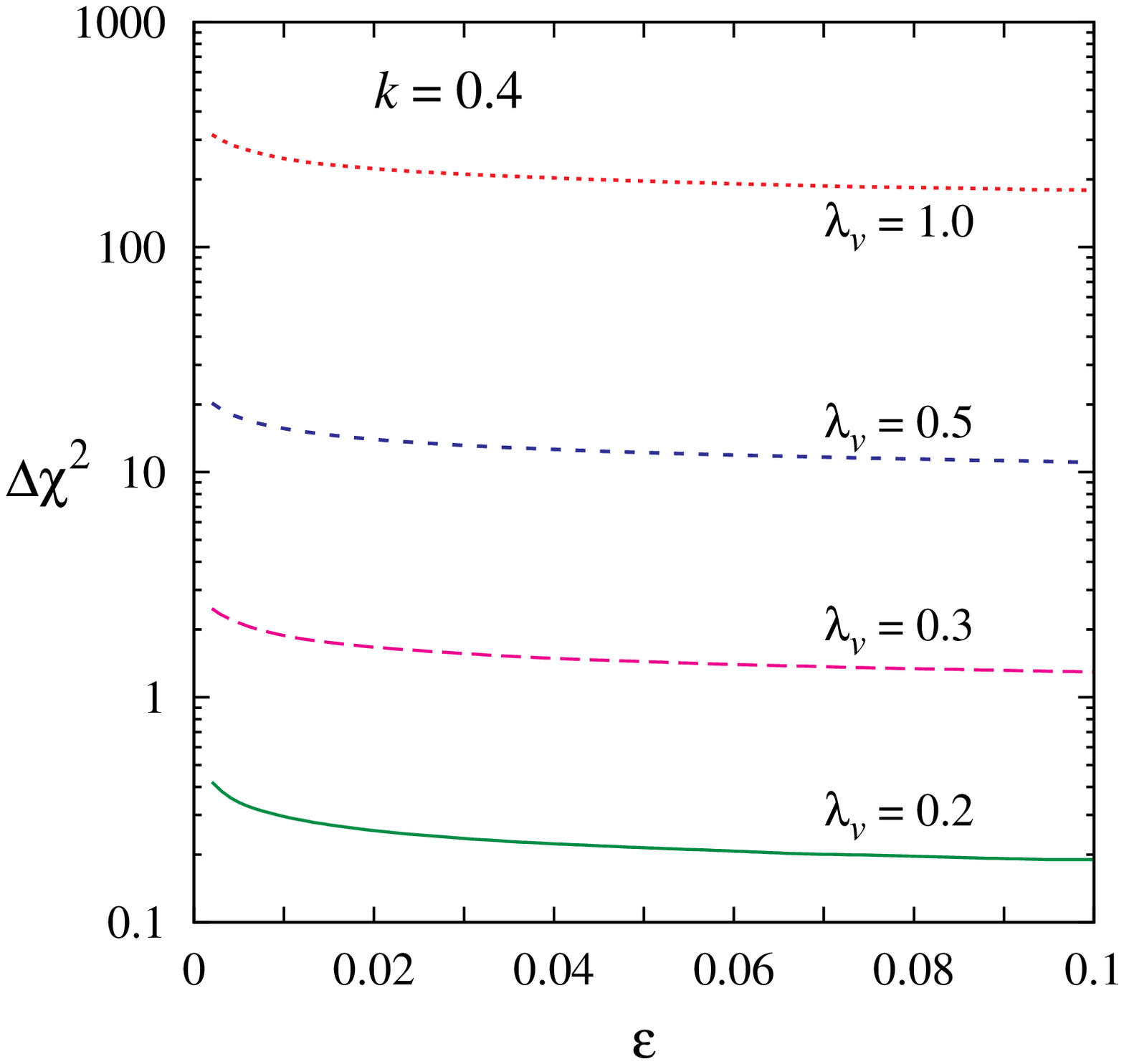}
\epsfxsize=7cm\epsfysize=9cm\epsfbox{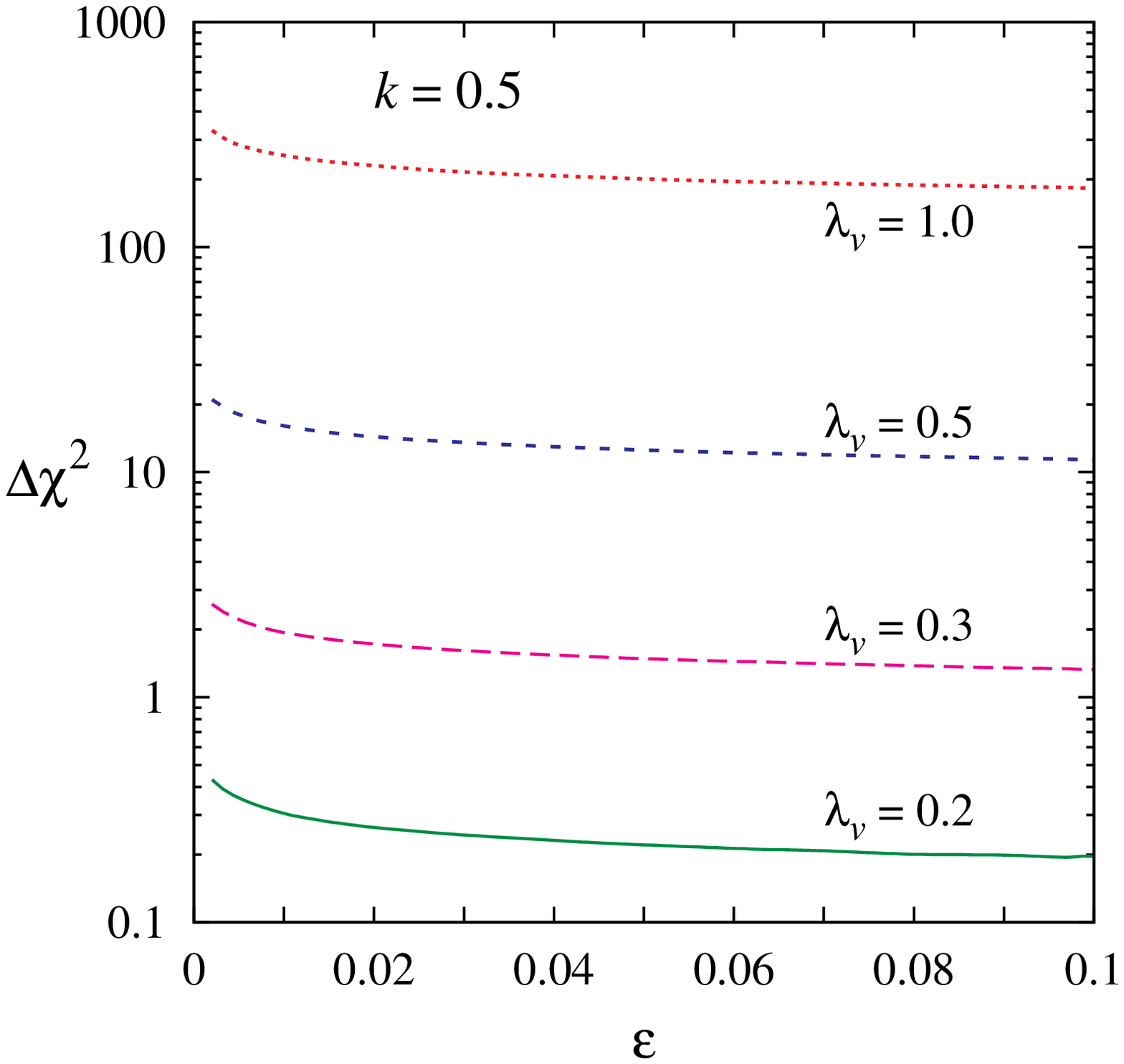}
}
\vspace*{-70pt}
\caption{\em The shift in the $\chi^2$ (as derived from the 22 observables 
listed in Ref.\cite{Csaki:2002gy}) as a function of $\epsilon$ for various
values of the parameter $\lambda_v$. The left and right panels refer 
to $k = 0.4$ and $k = 0.5$ respectively, while the parameter $\aleph$ 
has been fixed as in Fig.\ref{fig:epsi_aleph}.
}
\label{fig:epsi_chisq}
\end{figure}

\section{Beta function}
\label{sec:rg}
Grand unification remains a holy grail for scenarios of physics beyond
the SM, for not only does it provide a unification of forces, but also
a platform to answer questions pertaining to inflation and
baryogenesis on the one hand, and a formalism to understand fermion
masses on the other. Within the standard four-dimensional paradigm,
gauge coupling unification occurs, though only at scales in the
vicinity of $10^{15}$--$10^{16} \gev$, thereby putting a direct verification
of the paradigm beyond the reach of experiments in the foreseeable
future. A curious thing happens in the case of the universal
extra-dimensional scenarios. The renormalization group evolution of
the gauge couplings (which is logarithmic in the case of the SM) now
turns power-law\cite{Dienes:1998vg, Bhattacharyya:2006ym}. This can
be understood most easily in terms of the KK-reduction, whereby the
logarithmic contributions from each of the individual KK-excitations
sum up to give a power-law behaviour\footnote{Much the same would be
  seen if the entire calculation were to be done in the full
  five-dimensional theory. Care must be taken, though, in view of the
  inherently non-renormalizable nature of the theory.}.

For warped geometries, as is the case here, additional 
features arise. The absence of a KK-parity implies the existence 
of additional loops. A further complication is caused by the fact that,
owing to the nontrivial differences in their wavefunctions, distinct
KK-levels of the same field have differing coupling strengths. This already 
renders the evolution to be quite different from the UED case. 

And, finally, there is the issue of the graviton loops. Unlike in the
UED case, here the couplings of the KK-gravitons are non-negligible
and ought to be included.  On the other hand, such a inclusion cannot
be made in a straightforward fashion for the entire treatment of the
gravitation sector has been semi-classical and loop calculations with
gravitons are ill-defined.

In view of this, we desist from considering any graviton-loops.  This
can also be justified in the sense, that for a given KK-level, the
graviton is not only heavier than the SM excitations, but also has a
effective coupling\footnote{The graviton coupling is, of course, 
  dimensionful. What should be compared to the effective $\gym$ is
  the product of the graviton mass and its coupling. In the small $k$ 
  regime, this is indeed much smaller~\cite{Arun:2014dga,Arun:2015ubr}.}
 significantly smaller than them. Naively at least, the
graviton contributions to the gauge beta-functions would, thus, be
expected to be numerically small. Hence, while our results cannot be
termed exact, they are expected to be very good approximations of
calculations in the full theory.

Restricting ourselves to a discussion of the 
interactions between the SM fields, the one-loop $\beta$-functions can 
be calculated in a straightforward manner, considering the KK-excitations 
to be heavy particles with appropriate couplings (gauge, Yukawa) with 
the corresponding contributions to be included as a threshold is 
crossed\footnote{Since we are effecting only a one-loop calculations, 
neglecting the threshold effects is an excellent approximation.}.

As already mentioned, even this task is rendered difficult by the fact
that KK-number, or even KK-parity, is not conserved. However, as shown
in \cite{Arun:2015kva}, as far as the interactions of the zero mode
gauge fields are concerned, KK-number is indeed conserved. This is
exactly true as long as the modifications due to Higgs localization
can be neglected, which it can indeed be above the electroweak
symmetry breaking scale.  The change in evolution of the hypercharge
is straightforward as we only need to calculate the additional
contributions to the vacuum polarization. Similarly, for the
non-abelian component of the theory, the task, at one-loop order, is
easier for the triple-``gluon'' vertex\footnote{The same holds for the
  four-``gluon'' vertex as well, except that more diagrams need to be
  calculated.} (for all the vertices now respect KK-number
conservation) than for the gauge-fermion vertex. This can be
exploited, in conjunction with the appropriate Slavnov-Taylor identities (since
our six-dimensional Lagrangian is gauge invariant) to calculate the RG
flow for the other vertices as well.  Thus, the exercise is very
similar to that in the universal extra-dimension scenarios, but for
the added complication of unevenly placed KK-masses that need to be
calculated numerically\footnote{In actuality, after the first few
  levels, the rest can be rather well-fitted in terms of a bilinear
  function.}. It should be appreciated that the same results are
obtained for vertices that admit KK non-conservation, but only if all
the modes are taken into account.

\begin{figure}[!h]
\vspace*{-40pt}
\centerline{
\epsfxsize=5.75cm\epsfbox{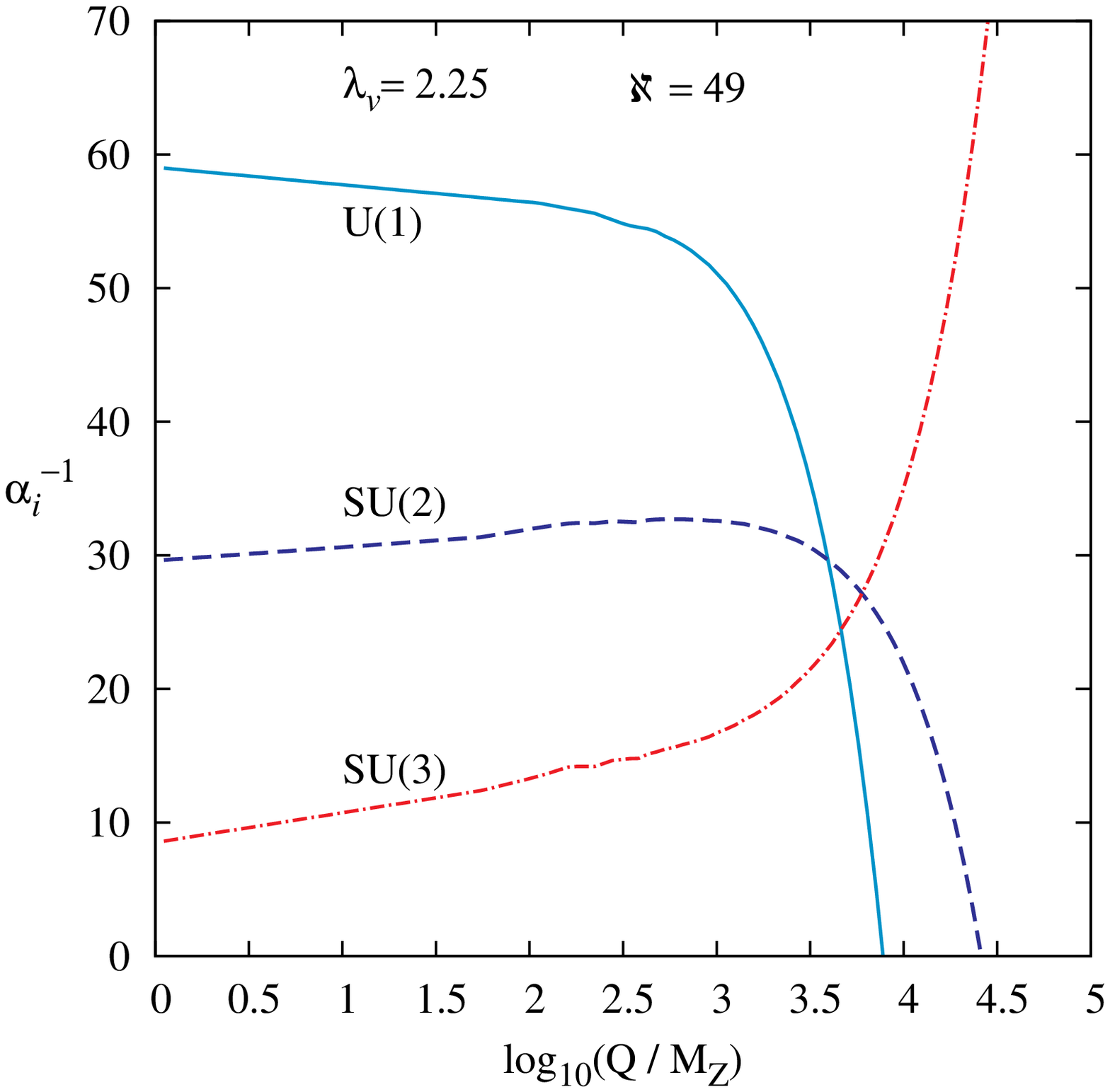}
\hspace*{-30pt}
\epsfxsize=5.75cm\epsfbox{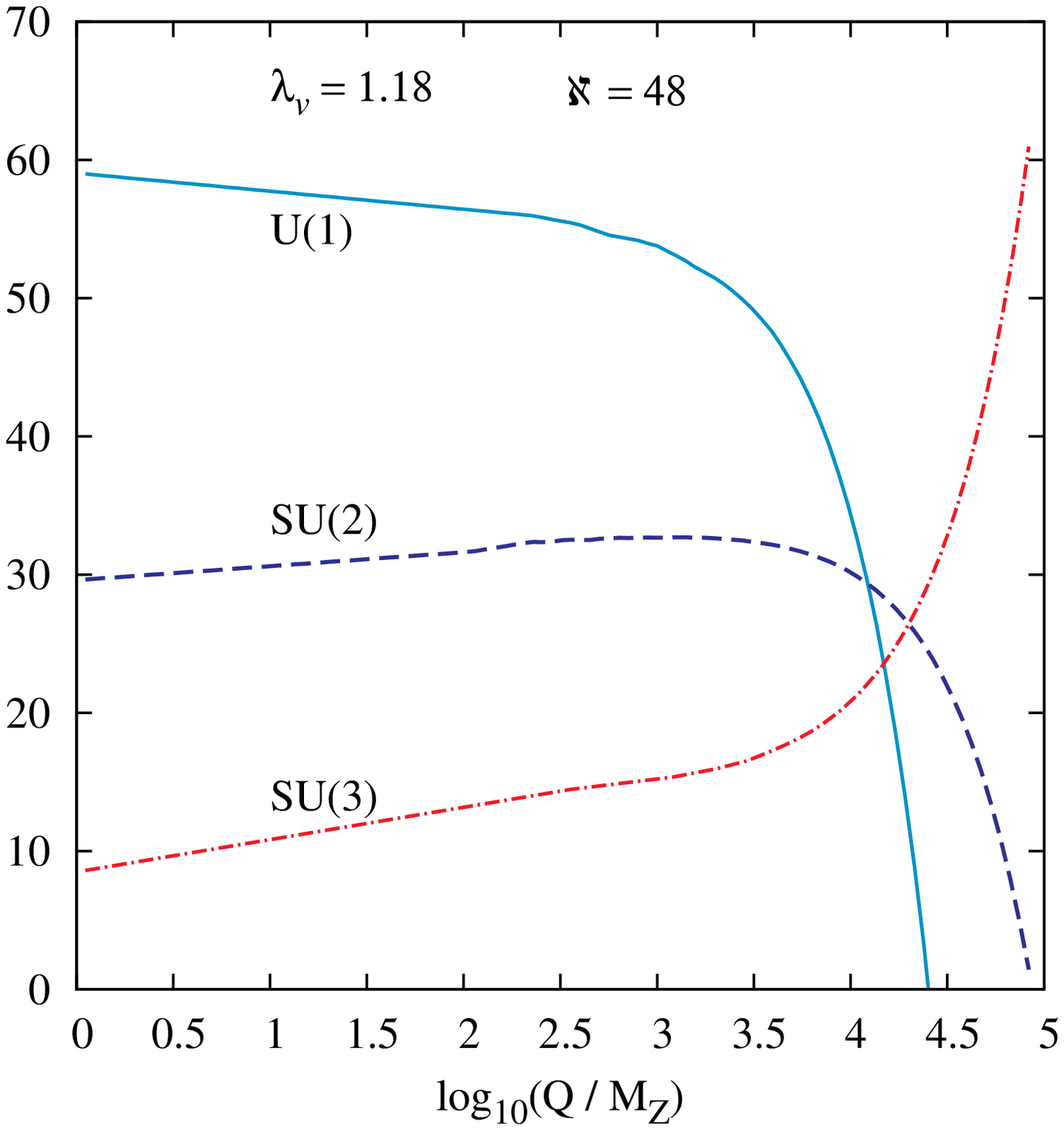}
\hspace*{-30pt}
\epsfxsize=5.75cm\epsfbox{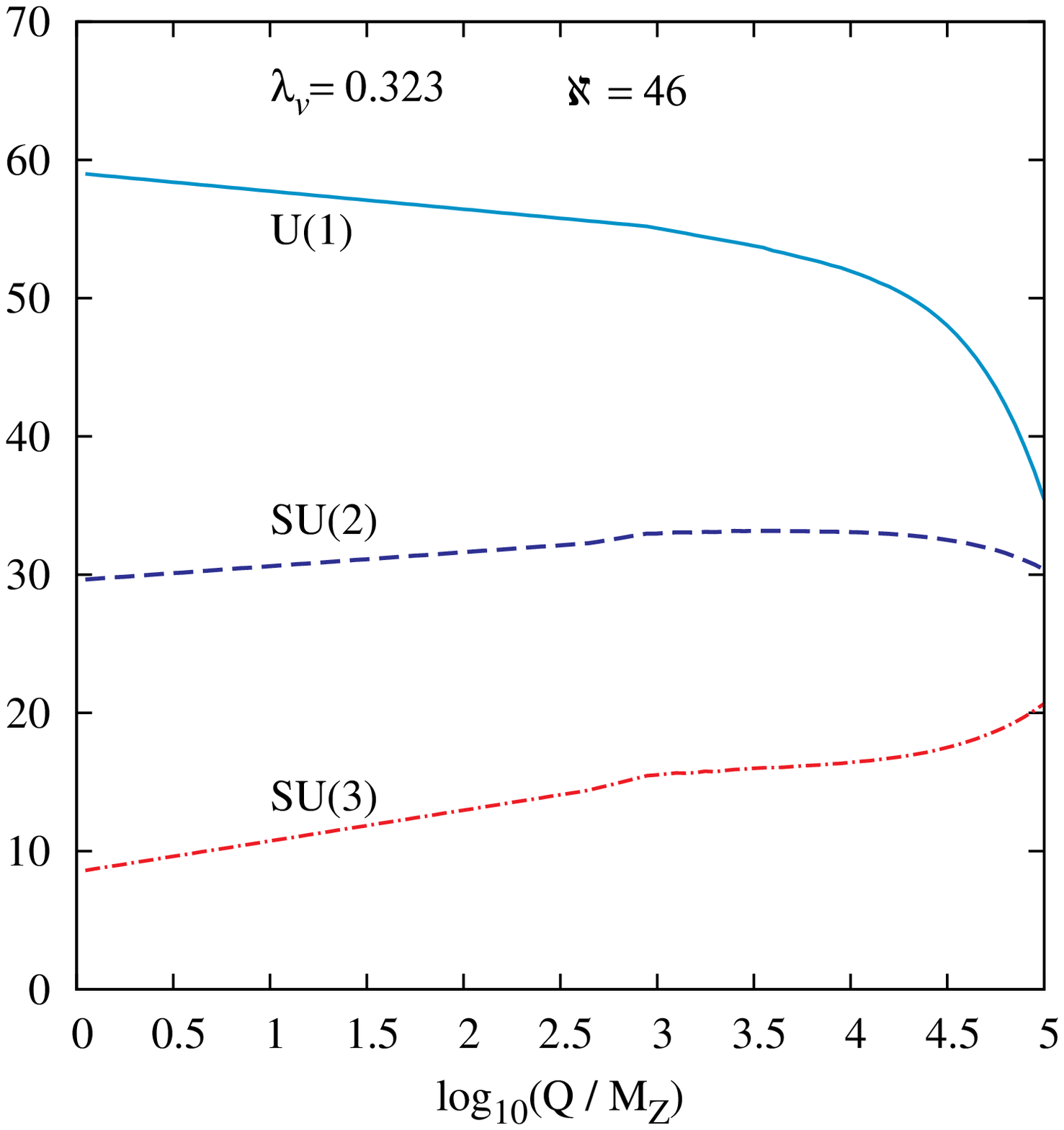}
}
\vspace*{-50pt}
\caption{\em Gauge coupling constant evolution for $k=0.5$, $\epsilon
  =0.1$. The three panels correspond to different $\lambda_v$ 
  values.}
\label{fig:running_coupling}
\end{figure}

In Fig.\ref{fig:running_coupling}, we display the evolution 
of the gauge coupling constants for a particular parameter point,
namely ($k=0.5, \epsilon = 0.1$) and some representative 
values of $\lambda_v$. As is expected, the evolution is indeed 
much faster than in the SM, and the ``unification'' scale is 
lowered to approximately $\approx 10^3$--$10^6 \tev$. The lower
$\lambda_v$ is, the higher are the masses for the KK-excitations, 
and, consequently, the higher is the unification scale. 
On the other hand, if we introduce a mechanism (such as those including
a custodial symmetry) that allows us to significantly lower 
the KK-masses, the unification scale would be lowered instead.

Two additional features are worth commenting on. The first is that,
not only the $U(1)_Y$ theory, but also the $SU(2)_L$ theory lacks
asymptotic freedom, a consequence of the number of new states in the
theory. This is quite analogous to the case of the
UED\cite{Bhattacharyya:2006ym}. A related feature is the presence of
some small kinks in the plots, visible most prominently for the case
of the $SU(3)$.  This, once agaian, is but reflective of momentary
change of the sign of the $\beta$-function and owes its origin to the
relative placements of the KK-excitations. Note that the latter feature is particularly sensitive to the order to which the
RG-equations are calculated, and stand to be significantly altered
once we go beyond treating thresholds as discrete steps. Furthermore, 
such effects could also play a role in resolving the lack of exact 
unification.

\section{Summary and outlook}
\label{sec:conclusion}
While a five-dimensional world with a warped metric and the SM fields
confined on a end-of-the-world brane (the RS scenario) offered a
tantalizing solution to the hierarchy problem, it suffers from the
obvious problem that no KK-excitation of the graviton has been
observed so far. Similarly, if one were to calculate amplitudes for
flavour-changing neutral currents, the low cut-off ($\sim 1\tev$) of
the theory implies that the dimension-six operators do not suffer a
large suppression and the resultant rates are too high. The first
problem can be solved~\cite{Arun:2014dga, Arun:2015ubr} courtesy
reduced couplings of the gravitons in a six-dimensional
generalization\cite{Choudhury:2006nj} of the original RS model with
nested warping. Indeed, the coupling can be suppressed well enough for
the recently reported diphoton excess at $750 \gev$ ~\cite{atlas_750,
  cms_750} to be explained in terms of such a resonance~\cite{Arun:2015ubr},
a feat impossible within the five-dimensional paradigm.

Allowing the fermions and gauge fields to propagate in the bulk is an
obvious antidote to the second problem, since four-Fermi operators are
now suppressed by higher powers of the ultraviolet cutoff. On the
other hand, doing so will bring into play KK-towers of the fermions
and gauge bosons and these, in turn, will effect low-energy
observables thereby inviting tight constraints from the indirect
measurement data obtained at LEP.  Indeed, within the five-dimensional
paradigm, such constraints push the gauge boson KK-masses well beyond
the reach of the
LHC~\cite{Rizzo:1999br,Chang:1999nh,Csaki:2002gy}. This brings back at
least a little hierarchy unless additional physics such as new
particles alongwith a custodial symmetry is invoked.

Clearly, both sets of problems could be addressed if one considers
bulk gauge bosons and fermions in a six-dimensional theory with nested
warpings, and the required formalism was introduced in Paper
I~\cite{Arun:2015kva}. Such a construction brings forth several
interesting consequences such as restrictions on the number of chiral
generations. Furthermore, with one particular tower disappearing
identically for each fermion species, if such KK-fermions (KK-bosons) can be produced 
at a collider, the signatures would be quite non-canonical.  
In the current paper,
we examine the issue of electroweak symmetry breaking in this scenario
as well as consider the phenomenological implications and constraints.

Contrary to the case of the fermions and gauge bosons, the Higgs
cannot percolate into the six-dimensional bulk, for it would bring
back the hierarchy problem. While it might seem that confining it to a
3-brane would be the simplest solution, this, unfortunately presents
some technical complications (as discussed in Paper I). 
Instead, we consider a novel mechanism
confining it to a 4-brane, with the Higgs acquiring a $x_5$--dependent
classical configuration thanks to an interplay between the potential
term and the nontrivial kinetic term endemic to a curved background.
The maximum $v$ of the classical configuration is naturally of the
order of the cutoff $R_y^{-1}$ (suffering only a mild suppression $0.1
\lapp \lambda_v \lapp 1$), but is warped down to the electroweak
scale. Interestingly, the simplest such construction puts a limit $k
\lapp 0.5$, beyond which tachyonic modes develop.  The
five-dimensional nature of the Higgs field is manifested in the shape
of KK-resonances, which often tend to be quite light if tree level
relations to equate the zero-mode mass to $125 \gev$.  However, once
quantum corrections are included, the KK-masses are lifted
considerably (alongwith significantly relaxing the constraint on the
parameter $k$). For example, for $k = 0.4$ and $\lambda_v = 0.5$, a
small perturbation $\beta_{-3} \approx 3.4 \times 10^{-4}$ leads to a
first excited mass $M_{h^{(1)}} \simeq 800 \gev$. This would be of
particular interest if the recently reported
excess~\cite{atlas_750,cms_750} in the diphoton channel is actually
confirmed. For while the graviton sector can also have such a
resonance~\cite{Arun:2015ubr}, allowing the SM fields into the bulk
not only forces us to a part of the parameter space 
that increases the mass of the first graviton resonance, but also
drastically suppresses thereby reducing the signal strength. Furthermore, if the resonance is
to be a graviton, then we should soon see excesses in other channels
as well (although the present data is inconclusive). On the other
hand, the aforementioned Higgs resonance would not decay to
$W/Z$-pairs through tree-level couplings, and with the couplings to
the top-sector also being modified considerably, it could present an
interesting alternative.

Of more immediate concern are the effects on low-energy phenomenology.
With flavour changing neutral current operators now being suppressed
by four powers of the UV-cutoff (in contrast with only three powers
for the analogous five-dimensional theory), the constraints from this
sector are minimal. On the other hand, the very confinement of the
Higgs onto a 4-brane introduces changes in the gauge-boson
wavefunctions that manifest themselves in the form of additional
tree-level contributions to the oblique parameters $S$ and
$T$. Similarly, the existence of the gauge-boson KK-resonances leads
to a change in $G_f$, the four-Fermi coupling. We perform a $\chi^2$
test using the data on 22 such precision-measured observables, to find
that the theory agrees very well with the low-energy data for
$\lambda_v \lapp 0.5$, and is virtually indistinguishable from the SM
for $\lambda_v \lapp 0.3$, both of which represent only a very small
hierarchy between the Higgs vev and the UV cutoff.

It is instructive to consider the reason for and the
  circumstances of this agreement.  For one, just as in the case of
the graviton-tower, the coupling of the gauge boson-tower with the SM
fermions are also somewhat suppressed.  The consequent reduction
  in $\delta G_f$ obviously helps. However, much of $\Delta\chi^2$
  accrues from the modification of the wave-function.  It is here
  that this scenario is not very different from the five-dimensional
  analogue. Consequently, the limits on the KK-masses are very
  similar, the present scenario doing only marginally better. What
is of more importance is that raising the KK-masses in the
five-dimensional theory begins to call into doubt the semi-classical
approximation that is the cornerstone of the treatment of the gravity
sector. Indeed, even with the introduction of additional physics and a
custodial symmetry, a fine tuning of ${\cal O}(10^{-2})$ would be
needed. In contrast, the six-dimensional theory studied here requires
only a small fine tuning of ${\cal O}(\lambda_v)$. Furthermore,
whereas the five-dimensional analogue would essentially push up all
the resonances (except, maybe, the radion) above the reach of the LHC,
this is not the case here. For, the Higgs resonances provide
additional handles that can the model can be probed with.  It would
also be interesting to examine this sector at the LHC, but we postpone
this to a later effort. 

And, finally, we come to the issue of RG evolution. With the
proliferation of states, naively, it would seem that the evolution
would be much faster than is the case for the five-dimensional theory.
This is not quite true, for a small $k$ means that the excitations in
the $x_5$-directions are typically heavier. However, certain features 
(such as the loss of asymptotic freedom for the $SU(2)$ interactions) 
are, understandably, quite similar to that in UED theories, but for 
the fact that the masses (and, hence, the thresholds) are non-uniformly 
spaced here. This, for example, leads to rapid changes in the sign of the 
beta-functions at certain intermediate points.

It might seem, overall, that by making the masses large, we 
have, essentially, decoupled the KK-sector. While this is forced 
upon us by the extremely good agreement of the low-energy observable
with the SM expectations, it should be realized that the required 
masses, apart from being somewhat lower than is the case for the 
five-dimensional theory, are perfectly commensurate with the 
applicability of the semi-classical treatment of the gravitational 
sector, and does not need the introduction of additional symmetries 
(as the RS case does) to bring down the scale. However, if such 
a custodial symmetry is indeed imposed, the mass scale can be brought 
down and interesting signals may be seen at the LHC itself. We 
leave this for a future study. Also postponed is a thorough 
investigation of the Higgs sector, especially the consequences
of our novel localization scheme wherein the scalar acquires a 
non-trivial classical configuration along a four-brane.


\section*{Acknowledgements}
MTA would like to thank UGC-CSIR, India for assistance under
Senior Research Fellowship Grant Sch/SRF/AA/139/F-123/2011-12. 
DC acknowledges partial support from the
European Union FP7 ITN INVISIBLES (Marie Curie Actions, PITN--GA--2011--289442),
and the Research and Development grant of the University of Delhi.

\end{document}